\documentclass[12pt]{article}
\usepackage{amsmath,amssymb,amsthm,amsxtra,overpic,bbm,bm,epsfig,ulem,multirow}
\usepackage{color,cite}
\usepackage{epstopdf}
\usepackage{booktabs}
\usepackage{footnote}
\usepackage{rotating}
\usepackage{hyperref}
\usepackage{url}
\usepackage{cite}
\usepackage{tikz}
\usetikzlibrary{arrows,shapes,snakes,automata,backgrounds,petri}
\usetikzlibrary{decorations.pathmorphing}
\usetikzlibrary{decorations.markings}
\usepackage{lineno}
\usepackage{subfigure}

\def\thefootnote{\fnsymbol{footnote}}

\usepackage{array}

\begin{document}

\vspace{0.2cm}

\begin{center}

{\Large\bf Atmospheric Neutrino Charged-Current Interactions at Large Liquid-Scintillator Detectors:} \\
\vspace{0.2cm}
{\Large\bf \underline{I. Physics of Neutrino-Antineutrino Discrimination}}

\end{center}

\vspace{0.2cm}

\begin{center}
{\bf Xinhai He~$^{a,b}$}~\footnote{Email: xhhe@ihep.ac.cn},
{\bf Gao-song Li~$^{a}$}~\footnote{Email: ligs@ihep.ac.cn},
{\bf Yu-Feng Li~$^{a,b}$}~\footnote{Email: liyufeng@ihep.ac.cn},
{\bf Wuming Luo~$^{a}$}~\footnote{Email: luowm@ihep.ac.cn},
{\bf Liang-jian Wen~$^{a}$}~\footnote{Email: wenlj@ihep.ac.cn},
 \\
\vspace{0.2cm}
{$^a$Institute of High Energy Physics, Chinese Academy of Sciences, Beijing 100049, China}\\
{$^b$School of Physical Sciences, University of Chinese Academy of Sciences, Beijing 100049, China}
\end{center}

\vspace{1.5cm}

\begin{abstract}
In this work, we present a systematic study of the event characteristics and physics of neutrino-antineutrino discrimination associated with atmospheric neutrino charged-current interactions in large liquid scintillator detectors. This study encompasses the primary neutrino interactions, the sequential second interactions of final-state particles, and the final neutron captures. We carefully investigate the properties of final-state charged leptons and hadrons, providing distinct distributions of inelasticity and captured neutron multiplicity for both neutrino and antineutrino interactions. These distributions are employed to assess the quantitative performance of neutrino-antineutrino discrimination. Our findings lay the groundwork for atmospheric neutrino oscillation studies in large liquid scintillator detectors, particularly in the determination of neutrino mass ordering.
\end{abstract}


\def\thefootnote{\arabic{footnote}}
\setcounter{footnote}{0}

\newpage

\section{Introduction}\label{sec:intro}
    
Over the past decades, the discovery of neutrino oscillation phenomena~\cite{Kajita:2016cak,McDonald:2016ixn} has proved the existence of neutrino masses and lepton flavor mixing.
In the standard three-flavor neutrino framework, the neutrino flavor eigenstates $\nu_\alpha$ ($\alpha=e,\mu,\tau$)
are expressed as linear superpositions of the neutrino mass eigenstates $\nu_i$ ($i=1,2,3$) through the lepton flavor mixing matrix~\cite{Pontecorvo:1957cp,Maki:1962mu,Pontecorvo:1967fh}, which is parametrized as
\begin{eqnarray}
U = \left( \begin{matrix}
 c_{12} c_{13} & s_{12} c_{13} & s_{13}e^{-{\rm i } \delta} \cr
 -s_{12} c_{23} - c_{12} s_{13} s_{23}e^{{\rm i } \delta} & 
 c_{12} c_{23} - s_{12} s_{13} s_{23}e^{{\rm i } \delta} & 
 c_{13} s_{23} \cr
 s_{12} s_{23} - c_{12} s_{13} c_{23} e^{{\rm i } \delta} & 
 -c_{12} s_{23} - s_{12} s_{13} c_{23}e^{{\rm i } \delta} & 
 c_{13} c_{23}
\cr \end{matrix} \right) ,
\label{eq:PMNS}
\end{eqnarray}
with $c_{ij} \equiv \cos\theta_{ij}$ and $s_{ij} \equiv \sin\theta_{ij}$ ($ij = 12,13,23$), and $\delta$ the Dirac CP-violating phase. Meanwhile, the frequencies of three neutrino oscillations are described by three mass-squared differences:
\begin{eqnarray}
\Delta m^2_{ij} = m_i^2 - m_j^2\;,
\label{eq:masss}
\end{eqnarray}
where $(i,j = 1,2,3,\, i>j)$ and $m_i$ is the mass of the $i$-th mass eigenstate $\nu_i$. 

In the standard three-neutrino oscillation framework, the mass-squared difference $\Delta m_{21}^2$ and the mixing angle $\theta_{12}$ have been determined through the KamLAND reactor antineutrino experiment and several solar neutrino experiments \cite{SNO:2011hxd}.
Meanwhile, the parameters $|\Delta m_{31}^2|$ and the mixing angle $\theta_{23}$ are primarily measured in atmospheric and accelerator neutrino experiments, including the T2K and Super-Kamiokande experiments \cite{T2K:2024wfn, Super-Kamiokande:2023ahc}. 
The third mixing angle, $\theta_{13}$, has been precisely measured in a series of reactor antineutrino experiments, particularly the Daya Bay experiment \cite{DayaBay:2012fng}, and later the long baseline accelerator neutrino experiments. 
While these parameters have been accurately determined, two significant unknowns remain in the current standard three-neutrino oscillation model: the neutrino mass ordering (NMO), which addresses the distinct patterns of neutrino mass spectrum, and lepton CP violation. Understanding these unknowns is crucial for advancing our knowledge of particle physics and the evolution of the universe.
    
As a fundamental property of neutrinos, NMO is a critical {parameter} for many other important measurements. On one hand, it influences the precision measurements of oscillation {parameters} and determination of lepton CP violation in neutrino oscillations. On the other hand, it provides constraints on determining the fundamental nature of neutrinos (i.e., Majorana or Dirac nature) and the observation of their absolute masses. The experimental determination of NMO can be categorized into two different methods. The first one is to use the Mikheev–Smirnov–Wolfenstein (MSW) matter effect~\cite{Wolfenstein:1977ue,Mikheyev:1985zog}, which is the foundation for long-baseline atmospheric and accelerator neutrino experiments, such as DUNE~\cite{DUNE:2021mtg}, Hyper-Kamiokande~\cite{Hyper-Kamiokande:2022smq}, PINGU~\cite{IceCube-PINGU:2014okk}, and ORCA~\cite{KM3NeT:2021ozk}. 
Due to the MSW effect, neutrinos in the energy range of 3 to 10 GeV will experience resonant flavor conversion in the neutrino interaction channel for normal mass ordering, while antineutrinos undergo the conversion in the inverted mass ordering case. 
This effect is especially pronounced in the electron flavor channels.
In accelerator neutrino experiments, the differentiation between neutrinos and antineutrinos can be accomplished by adjusting the configuration of the accelerator beam. Conversely, in the atmospheric neutrino experiments, the simultaneous presence of both neutrino and antineutrino types necessitates a clear distinction between them, which is essential for accurately determining the mass ordering.

The Jiangmen Underground Neutrino Observatory (JUNO)~\cite{JUNO:2015zny,JUNO:2024jaw}, with a baseline of 52.5 km and a large liquid scintillator (LS) detector of 20 kton, adopts a unique approach by leveraging the interference effects of quasi-vacuum oscillation of reactor antineutrinos, setting it apart from other neutrino oscillation experiments. Meanwhile, JUNO is also capable of detecting the natural atmospheric neutrinos, and the combination of reactor and atmospheric neutrino measurements is expected to enhance the sensitivity to the NMO. To achieve this goal, robust capabilities in neutrino directionality and type discrimination are essential. In Refs.~\cite{Yang:2023rbg,Liu:2025fry}, a preliminary attempt using machine learning techniques has been developed. However, the underlying mechanisms and physical origins remain to be clearly understood.   

In this work, we investigate the properties of event characteristics of GeV atmospheric charged current (CC) neutrino interactions in LS detectors and focus on the corresponding neutrino-antineutrino identification capabilities. 
Our study delves into fundamental physics involved in identifying atmospheric neutrinos within large LS detectors through comprehensive Monte Carlo simulations. Due to high light yields and low energy threshold, these detectors are particularly effective at capturing final-state hadronic components, which provide a distinctive statistical approach to differentiate between neutrinos and antineutrinos.
Neutron tagging has proven to be particularly effective in {LS detectors}, in which the neutron multiplicity resulting from atmospheric neutrino interactions serves as a key factor for {neutrino-antineutrino discrimination}. The inelasticity of neutrino interactions, defined as the ratio of the energy carried by final-state hadrons to the total energy transfer, is another important parameter that contributes to this differentiation.
Finally, we introduce a boosted decision tree (BDT) method to demonstrate the effectiveness of these two physical parameters in the neutrino-antineutrino discrimination. 

The remainder of this paper is organized as follows. In Sec.~\ref{sec:simulation}, we will introduce the interaction tools used to generate final-state particles and perform detector simulations for this analysis. Sec.~\ref{sec:pid} will focus on the characteristics of both final-state leptons and hadrons in the context of detector response and physics. The effects of the detector size on the characteristics of final-state particles will be discussed in Sec.~\ref{sec:geometry}. Finally, the performance of neutrino and antineutrino discrimination using BDT method will be presented in Sec.~\ref{sec:separation}.

\section{Framework of simulation calculations}\label{sec:simulation}

\begin{figure}
        \centering
        \includegraphics[width=0.68\textwidth]{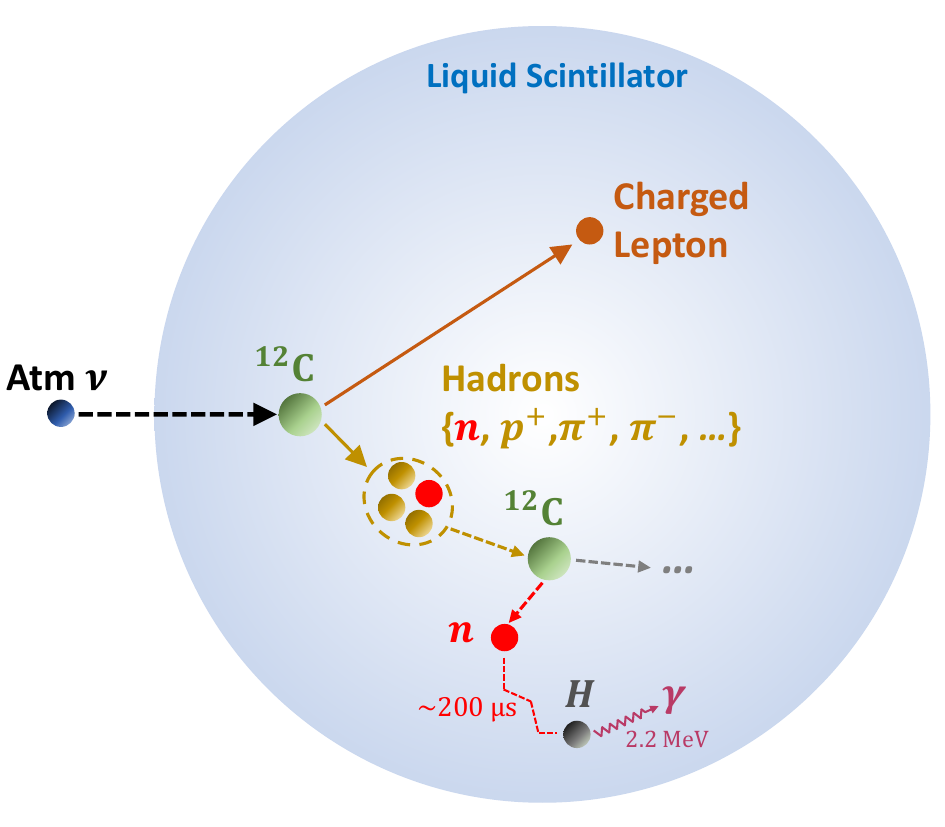}
        \caption{The schematic diagram for the atmospheric neutrino interactions in LS, including the charged current interactions, the sequential secondary interactions, and final neutron captures.}
        \label{fig:neutrino_interaction}
    \end{figure}

For this study, 
the simulation of neutrino CC interactions with $^{12}$C ($\sim$88\%) and $^{1}$H ($\sim$12\%) in the LS detectors is carried out by using the GENIE generator (Version 3.2.0) 
in which the dominant contribution to the cross section comes {mainly} from quasi-elastic scattering (QEL), nuclear resonance production (RES) and deep inelastic scattering (DIS) across different energy ranges. The secondary interactions of final-state particles in LS are also considered using the Geant4 simulation (Version 10.4.2).
In this calculation, four different neutrino types ($\nu_{e}$, $\overline{\nu}_{e}$, $\nu_{\mu}$, $\overline{\nu}_{\mu}$) with one million events for each neutrino type sampled with isotropic directions are input to GENIE. The tune of G18\_10b\_02\_11b is employed in this study, and a detailed description of the tuned model in GENIE V3 can be found in Ref.~\cite{GENIE:2021zuu}. {The neutrino fluxes used as input for GENIE are adjusted to be uniform in order to eliminate energy dependence.}
The final-state particles generated by GENIE are injected into a standalone Geant4 framework to simulate their secondary interactions in LS. {For each category of neutrino interaction events, the events are sampled with a uniform vertex distribution within the LS volume.} A customized Livermore physics model, which includes the positronium process, is used to simulate electromagnetic interactions~\cite{refId0}. Additionally, a QGSP\_BERT\_HP-based model with a modified neutron capture process is employed to simulate the secondary hadronic interactions~\cite{Allison2016}.
{As an illustrative framework, a simplified JUNO-like detector has been constructed in Geant4 to incorporate the primary features of the LS medium and basic detector geometries~\cite{JUNO:2015zny, Lin:2022htc}. This model includes 20 kton of LS contained within an acrylic sphere with a radius of 17.7 m. A total of 17,612 20-inch photomultiplier tubes (PMTs) are uniformly installed on the spherical surface of a stainless steel structure with a radius of 20 m. The space between the acrylic sphere and the PMTs is filled with purified water. No other detector components are included in this simplified framework. Additionally, in the simulation, we directly model the geometries of the PMTs and record the number of incident optical photons on each PMT. However, detailed simulations of the front-end electronics, as well as the subsequent waveform and event reconstruction processes, are not considered in this work. Correspondingly, no energy threshold is taken into account.}
A schematic diagram is illustrated in Figure~\ref{fig:neutrino_interaction} for the simulation of atmospheric neutrino interactions in LS, including the atmospheric neutrino interactions, the production of charged leptons and final-state hadrons, the sequential second interactions, and final neutron captures on {hydrogen}.

\begin{figure}
        \centering
        \subfigure[]{
            \includegraphics[width=0.48\textwidth]{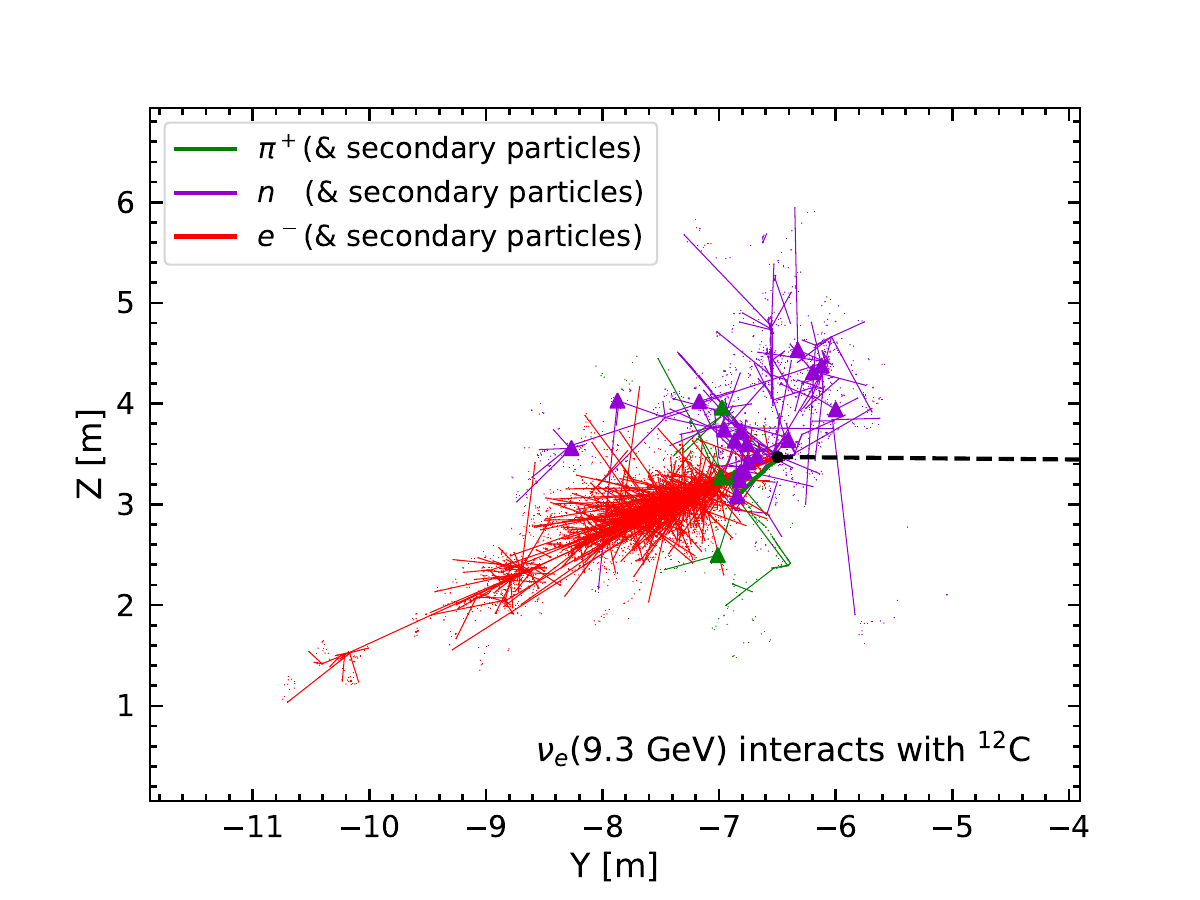}
            \label{fig:interaction_nu_e}
        }
        \hfill
        \subfigure[]{
            \includegraphics[width=0.48\textwidth]{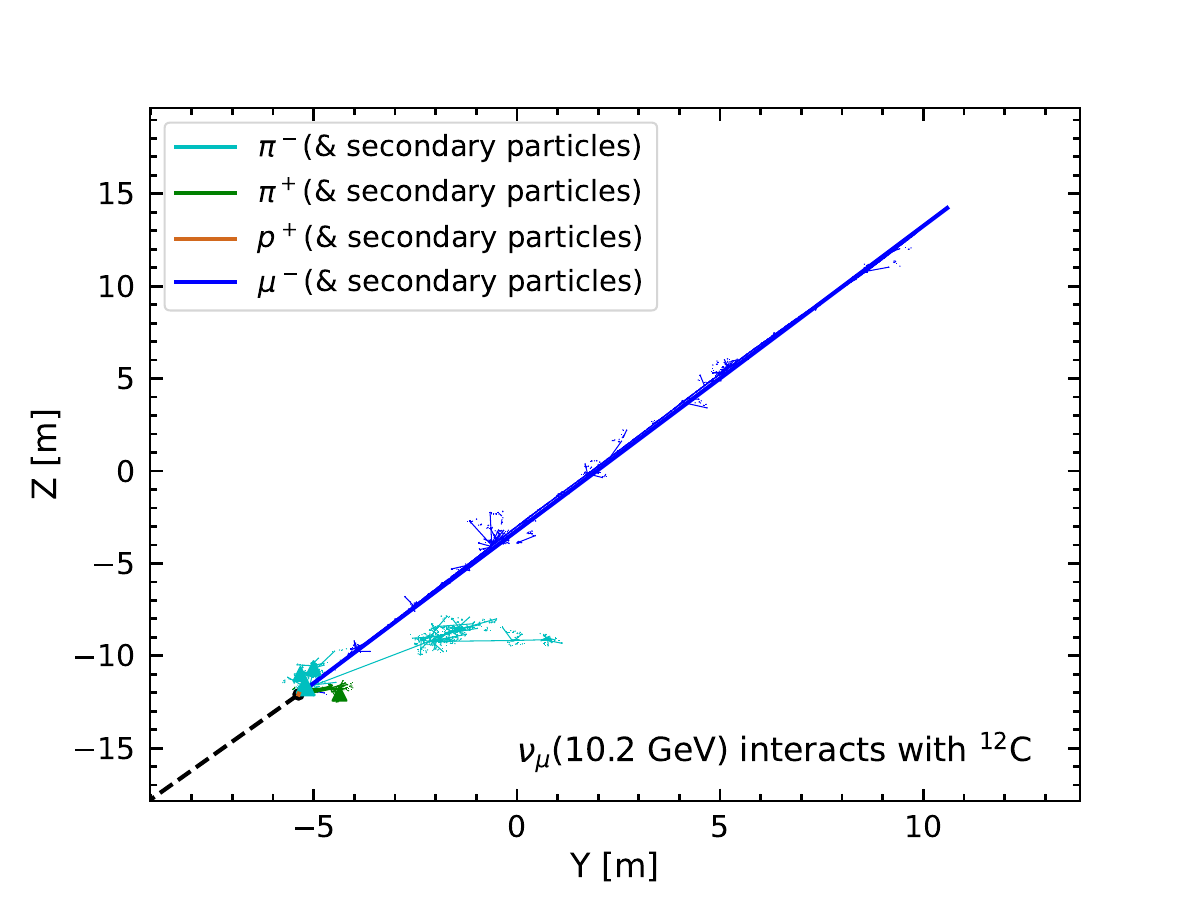}
            \label{fig:interaction_nu_mu}
        }
        \caption{Typical examples of charged current interaction between $\nu_e$ (left panel) or $\nu_\mu$ (right panel) and $\rm{^{12}C}$. The solid lines show the tracks of final-state particles and their secondary particles, and the dashed shows the incoming atmospheric neutrinos. The triangles represent the positions of neutrons capture.}
        \label{fig:interaction}
    \end{figure}

In Figure~\ref{fig:interaction}, we illustrate two distinct examples of charged current interactions of atmospheric neutrinos, each with an energy of approximately 10 GeV. The solid lines depict the tracks of the final-state particles and their resulting secondary particles, while the dashed lines represent the incoming atmospheric neutrinos. The triangles indicate the locations where neutron captures occur. 
In the case of electron neutrinos, the resulting final-state electron initiates an electromagnetic shower, which subsequently produces secondary electrons and gamma rays. Concurrently, the final-state hadronic shower resulting from $\nu_e$ interactions is also generated, introducing additional energy deposition and the possible production of neutrons or other hadronic particles. For muon neutrinos, the final-state muon deposits energy primarily through ionization, leaving a discernible track within the detector medium. It is important to note that in this scenario, the muon track is considerably easier to distinguish from the hadronic shower.
    
\section{Characteristics of Charged Current Interactions}
\label{sec:pid}

In this section, we present detailed event characteristics of atmospheric neutrino interactions using the simulated data samples. For the charged leptons, the distinct features of electrons and muons are manifested as typical electromagnetic showers and tracks, respectively. The resulting topological time and spatial patterns of detected photon electrons in the PMTs can be utilized for the flavor identification.
Meanwhile, the detection of hadronic components mostly contributes to the neutrino and antineutrino discrimination.
The inelasticity and the neutron multiplicity are two essential parameters for this purpose, and will be discussed in this section. These parameters provide valuable insights into the dynamics of the neutrino interactions and second interactions occurring within the LS, and thus play a crucial role in distinguishing between neutrinos and antineutrinos.

\subsection{Topology of Lepton and Hadronic Components}
\label{section:lepton}

When an electron interacts within the LS, it initiates an electromagnetic shower—a cascade of secondary particles, including gamma rays and additional electrons. This process produces abundant scintillation light over an extended area, resulting in a diffuse spatial distribution of light. In contrast, a traversing muon primarily ionizes the medium, depositing energy in a more localized manner along a straight, narrow track, which generates a brighter and more concentrated scintillation signal.
Hadronic components, on the other hand, are characterized by complex, clustered patterns in their showers due to interactions with multiple secondary particles in the LS. Consequently, PMTs record a broad, multi-point pattern from both electron-induced and hadronic showers, while muons produce a well-defined linear track. The distinct spatio-temporal profiles of these light patterns, as reflected in the PMT hit-time distributions, provide a robust basis for particle identification and neutrino interaction discrimination in the LS medium.
In addition to these primary light patterns, the delayed signatures of neutrino interactions, such as neutron capture and delayed beta decays of nuclei, offer further information for identifying neutrino interactions.

    \begin{figure}
        \centering
        \subfigure[]{
            \includegraphics[width=1\textwidth]{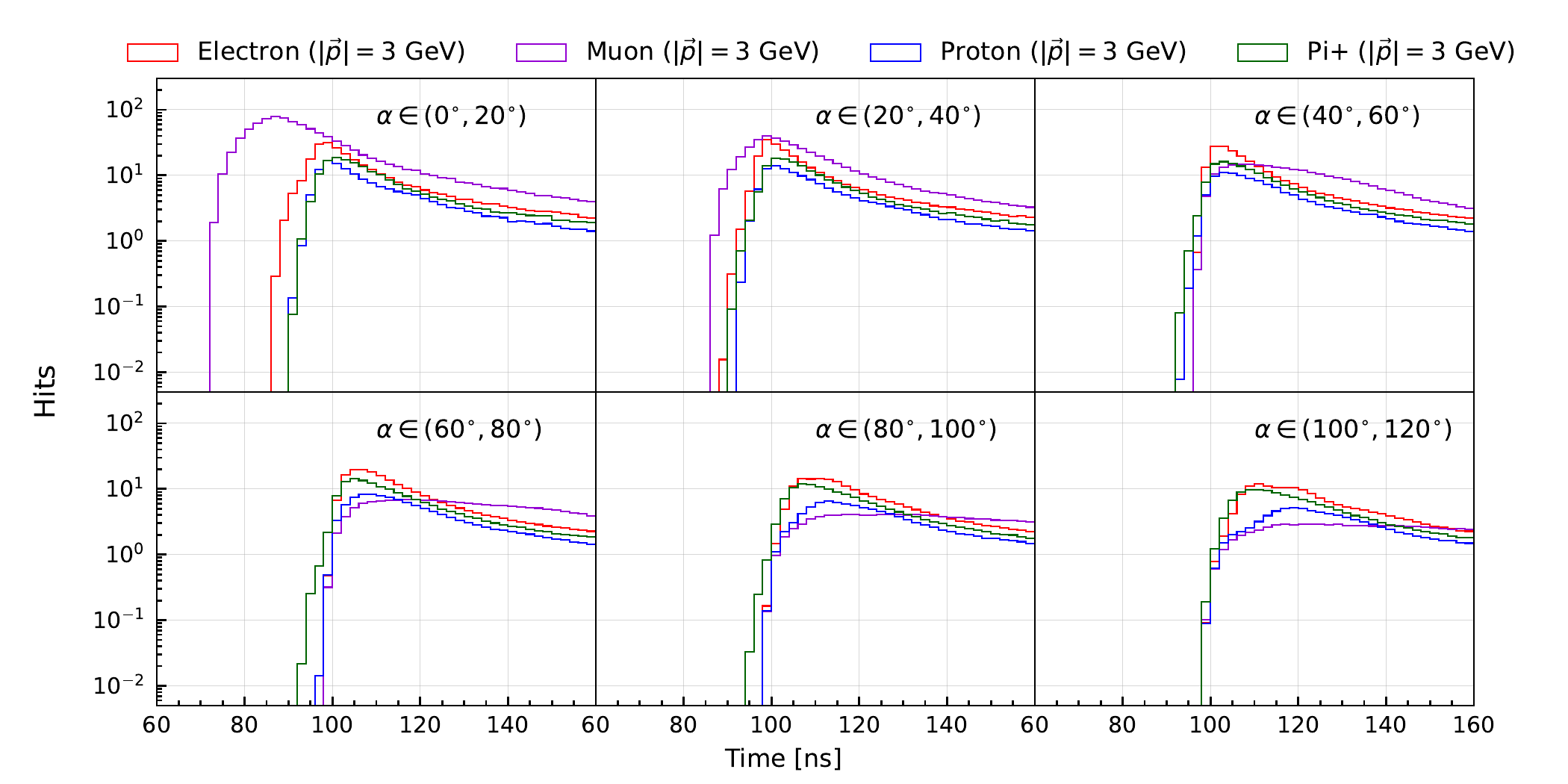}
        }
        \hfill
        \subfigure[]{
            \includegraphics[width=1\textwidth]{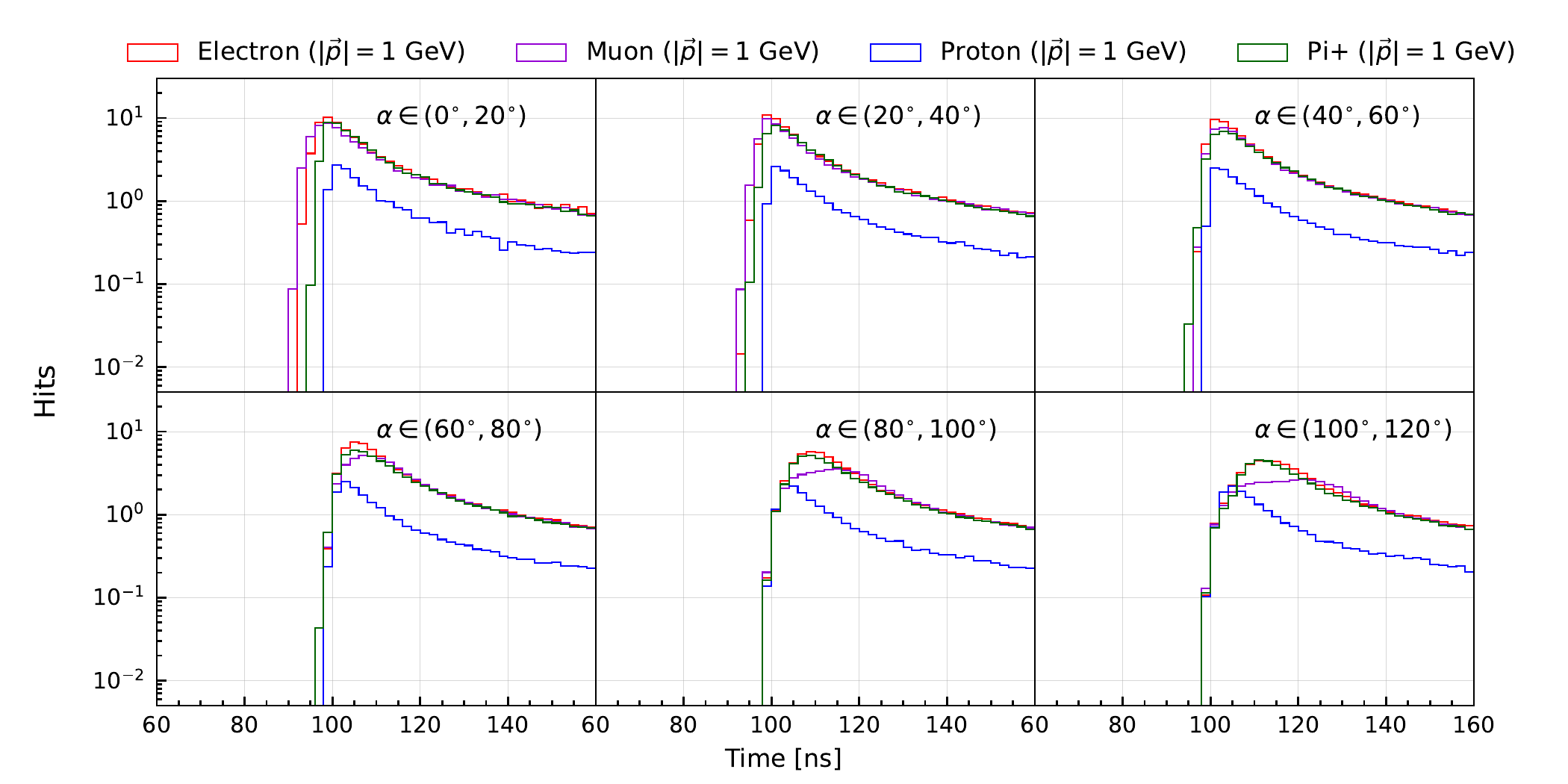}
        }
        \caption{
        {PMT hit-time distributions in different angular regions ($\alpha$), where $\alpha$ is the angle between the PMT position and the charged-particle momentum. Lines show electrons, muons, protons and pions with momenta of at 3\,GeV (a) and 1\,GeV (b). In the GEANT4 simulation, secondary interactions and products of these particles are enabled.}}
        \label{fig:particle_hittime}
    \end{figure}

Figure~\ref{fig:particle_hittime} illustrates the distinct signatures of hit-time distributions for electrons, muons, protons and pions across different angular regions, where $\alpha$ represents the angle between the incoming lepton momentum direction and the PMT position. Colored lines correspond to electrons, muons, protons and pions with momenta of 3\,GeV (upper panels) and 1\,GeV (lower panels), respectively. For each panel, the plots are arranged with increasing $\alpha$ values from upper left to lower right. Note that all the particles originate at the detector centre and the distributions are averaged and normalized over PMTs within each selected $\alpha$ region. 
From the figure, we observe that at multi-GeV energies, the hit-time profiles of muons and electrons differ distinctly. Higher momentum shifts the first hits to earlier times---a trend more pronounced for muons than for electrons, and stronger at smaller $\alpha$ values. Moreover, 3\,GeV muons exhibit more late-time hits than electrons of the same momentum at larger $\alpha$, an effect that diminishes as the momentum decreases.
Regarding hadrons, protons and $\pi^{+}$ {mesons} with higher momentum (3\,GeV) induce secondary interaction processes, such as inelastic scattering or decays, which generate secondary particles that contribute to hadronic showers. As a result, PMTs near these secondary particles may register signals much earlier, as illustrated in Figure~\ref{fig:particle_hittime}. Unlike the charged leptons and $\pi^{+}$'s, protons produce fewer photons due to {stronger quenching effects}. Additionally, at lower momenta, protons can be considered as a single point-like source of light.

To quantify the different patterns between the signature of the aforementioned lepton and hadrons, three PMT-level observables are introduced to capture their distinct hit patterns~\cite{Liu:2025fry}:
\begin{itemize}
\item \textbf{FHT}: the hit time of the first photoelectron collected by the PMT.
\item \textbf{NPE}: the total number of photoelectrons collected by the PMT.
\item \textbf{Slope}: the slope of the rising edge of the PMT hit distribution, defined using the interval between the FHT and the peak time.
\end{itemize}

Notably, for the LS detectors that utilize the PMTs as detection units, custom Flash Analog-to-Digital Converters are widely employed in the electronic systems~\cite{Coppi:2023nlv}. This design enables accurate acquisition of all three PMT-level observables in realistic detection scenarios.

    \begin{figure}
        \centering
        \includegraphics[width=0.95\textwidth]{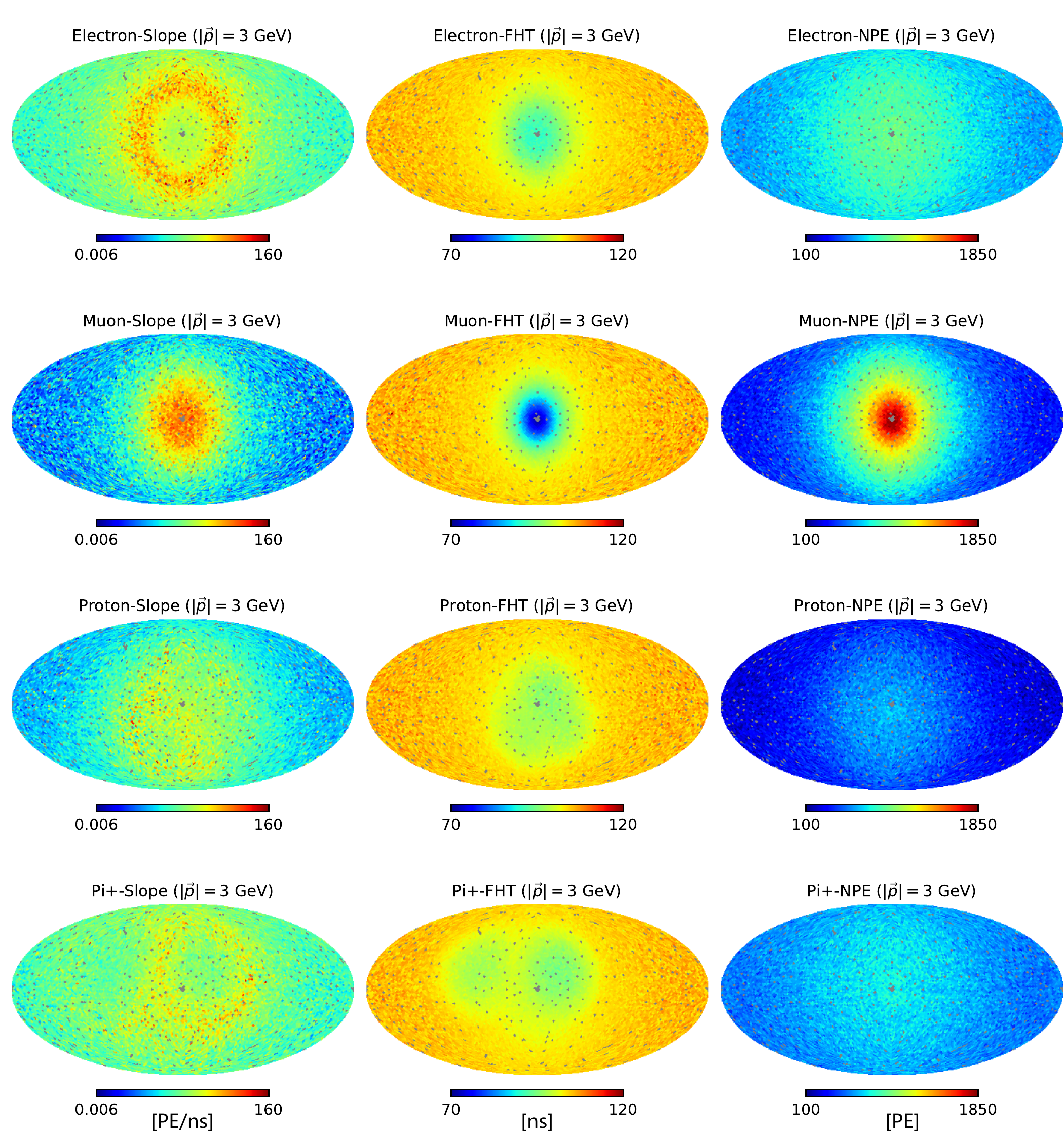}
        \caption{
        Mollweide projections of the PMT-level observables for electrons, muons, protons and pions with 3\,GeV momentum. Columns from left to right show Slope, FHT and NPE. All particles originate perpendicular to the plane, pointing outward. In the GEANT4 simulation, secondary interactions and products of these particles are enabled.}
    \label{fig:moll_pattern_3GeV}
    \end{figure}

    \begin{figure}
        \centering
        \includegraphics[width=0.95\textwidth]{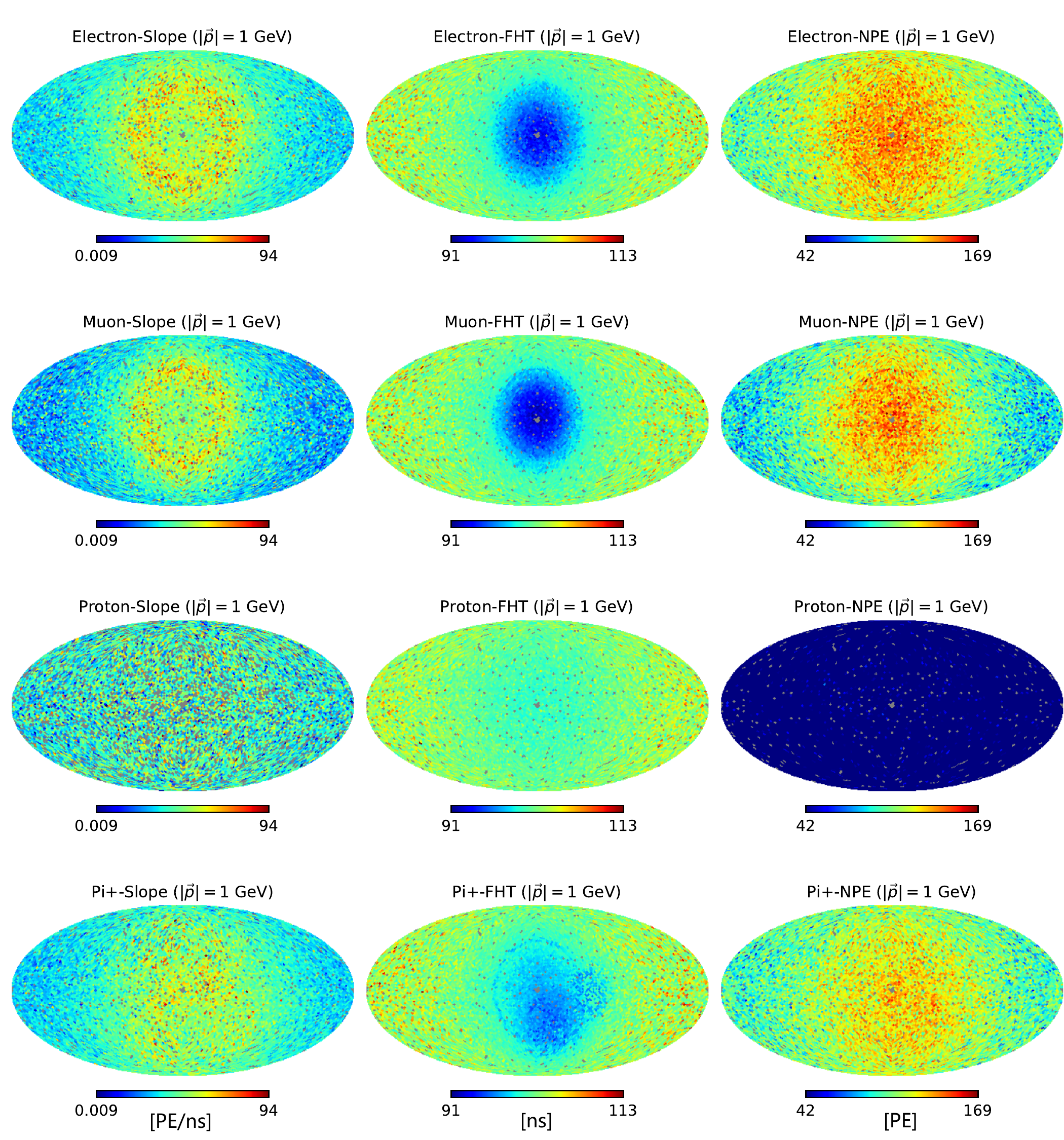}
        \caption{
        Mollweide projections of the PMT-level observables for electrons, muons, protons and pions with 1\,GeV momentum. Columns from left to right show Slope, FHT and NPE. All particles originate perpendicular to the plane, pointing outward. In the GEANT4 simulation, secondary interactions and products of these particles are enabled.}
       \label{fig:moll_pattern_1GeV}
    \end{figure}

For illustration, these PMT observables are quantitatively depicted in Figure~\ref{fig:moll_pattern_3GeV} and Figure~\ref{fig:moll_pattern_1GeV} using the Mollweide projection. This equal-area pseudo-cylindrical map projection preserves the relative area of all regions while representing the spherical surface on a flat plane. The left, middle, and right panels correspond to the Slope, FHT, and NPE variables, respectively. The momenta of the injected particles in Figure~\ref{fig:moll_pattern_3GeV} and Figure~\ref{fig:moll_pattern_1GeV} are 3\,GeV and 1\,GeV, respectively, with the momentum directions being perpendicular to the plane and pointing outward. 

Compared to electrons, a 3\,GeV muon can travel roughly 15\,m in LS, comparable to the $\sim$20\,m radius of the stainless-steel structure of JUNO that houses the inward-facing PMTs. When the muon stops near the PMT array, PMTs aligned with its trajectory receive a bright, prompt light signal. This yields a distinct slope pattern (left column of Figure~\ref{fig:moll_pattern_3GeV}) and a rapid fall-off of NPE with the increasing $\alpha$ angle relative to the muon direction. These features allow an excellent discrimination of electron and muon neutrinos. 
At 1\,GeV, however, the muon track shortens, and the detector response behaves more like a point-like source. The hit-time distributions of 1 GeV muons are similar to those of electrons, resulting in less distinctive PMT-level signatures (Figure~\ref{fig:moll_pattern_1GeV}). Consequently, flavor-identification performance is expected to degrade at lower neutrino energies as shown in Ref.~\cite{Liu:2025fry}.

The Mollweide projection of the hadrons can be understood through the distinct patterns of multiple hadronic showers, which exhibit clustered structures that differ from those produced by electrons and muons. These clusters arise from ionization-induced scintillation in the LS caused by multiple secondary particles.
When the momentum decreases to 1 GeV, protons deposit energy almost exclusively through ionization. Due to their larger mass and greater ionization energy loss, they leave very short tracks in the LS, resulting in a point-like light source at the center of the detector. Consequently, Figure~\ref{fig:moll_pattern_1GeV} shows relatively small variations in the number of photons received by the PMTs at different positions. In contrast, for $\pi^{+}$ particles, in addition to energy deposition via their own ionization, the charged leptons produced from their decay also contribute to ionization-induced scintillation. Therefore, even at lower energies, the PMT hit patterns generated by pions still exhibit the pattern of multiple clustered structures.

From the discussion above, it is evident that different charged leptons and hadrons elicit varying responses in the LS detector. This variability enables the discrimination between charged-current and neutral-current interactions, as well as flavor identification. {These PMT-level observables constitute a two-dimensional data table for each physics event, which can be utilized in deep learning models that employ multi-layer neural networks to perform effective feature extraction and reduce the dimensionality of the data.
In Ref.~\cite{Liu:2025fry}, it has been demonstrated that excellent discrimination performance can be achieved using {point-cloud based} machine learning methods under conditions that closely resemble actual detector operations. This makes the separation of neutrinos and antineutrinos feasible in realistic experimental observations.
}

\subsection{Inelasticity and Neutron Multiplicity}
\label{section:inelacticity}

With the fundamental patterns of electromagnetic showers, tracks, and multiple hadronic showers for electrons, muons, and hadronic components established, we are now prepared to discuss the parameters that can be utilized for neutrino and antineutrino discrimination. Generally, the CC interactions of the neutrino and nucleus can be described in a semi-inclusive manner as:
    \begin{equation}
        \nu + N \rightarrow M + l^{\pm} + X\,,
    \label{eq:CC_reaction}
    \end{equation}
where $N$ denotes the initial nucleus, $M$ the daughter nucleus, $l^{\pm}$ the outgoing charged lepton, and $X$ the hadronic final state. The inelasticity $y$ for such {interactions} is defined as
    \begin{equation}
        y = \frac{E_{\nu}-E_{l}}{E_{\nu}}\;,
    \label{eq:inelasticity}
    \end{equation}
with $E_{\nu}$ and $E_{l}$ the energies of the incoming neutrino and outgoing lepton, respectively.

\begin{figure}
        \centering
        \includegraphics[width=1\textwidth]{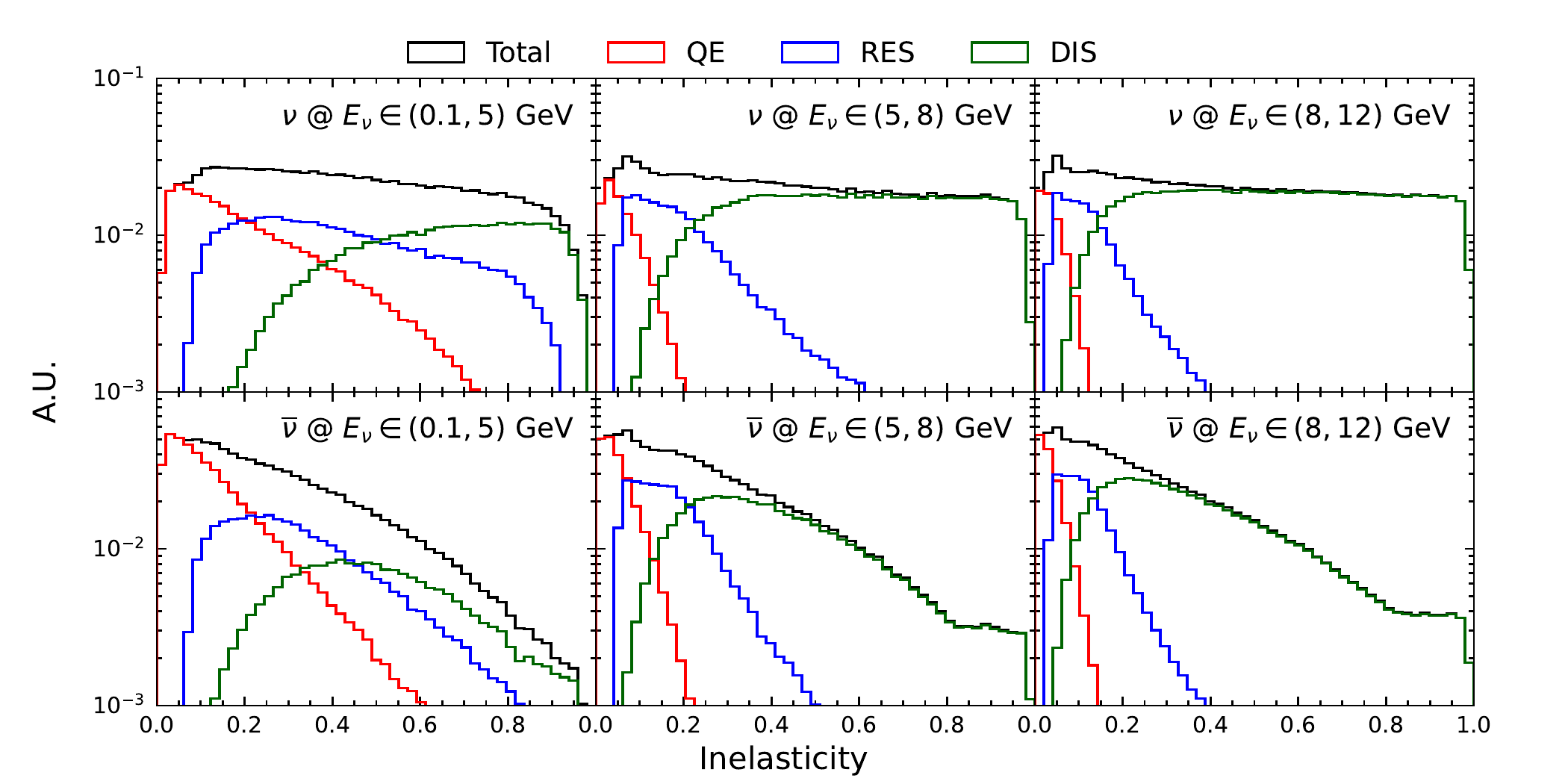}
        \caption{The inelasticity distributions for neutrino interactions (upper panel) and antineutrino interactions (lower panel) are presented across various energy ranges. For comparison, the individual distributions of QEL, RES, and DIS are also illustrated. Each distribution is normalized by the number of events within each energy range.}
        \label{fig:inelasticity}
    \end{figure}

The CC interactions of neutrinos and antineutrinos exhibit distinct inelasticity distributions (Figure~\ref{fig:inelasticity}), owing to differences in their chirality and to the parton distributions within nucleons. Neutrinos tend to transfer a larger fraction of their energy to the hadronic system than antineutrinos, leading to broader hadronic energy sharing. At low energies, quasi-elastic (QEL) scattering dominates, characterized by small $y$ values; here the neutrino interacts with a single nucleon, leaving most of its energy with the outgoing lepton. 

As the neutrino energy rises, deep inelastic scattering (DIS) becomes prevalent where neutrinos scatter off quarks within nucleons, producing larger $y$ values and a more diversified hadronic final state. The figure also shows the separate contributions from QEL, resonance (RES), and DIS processes, each normalized per energy bin. A clear asymmetry is observed between neutrinos and antineutrinos: antineutrinos systematically transfer less energy to hadrons. This difference provides a key handle for distinguishing the two interaction types.

Recent studies have demonstrated the reconstruction of the inelasticity parameter in water or ice Cherenkov detectors (Ref.~\cite{IceCube:2018pgc}). This approach enhances the separation of neutrinos and antineutrinos and improves sensitivity to the NMO (Refs.~\cite{Ribordy:2013xea, Olavarrieta:2024eaq}). In this work, LS detectors, with their lower energy thresholds, offer distinct advantages for detecting the hadronic components of neutrino interactions. Consequently, improved reconstruction of the inelasticity parameter is anticipated in LS-based detection, with detailed results to be reported in a separate publication.

    \begin{figure}
        \centering
        \includegraphics[width=0.95\textwidth]{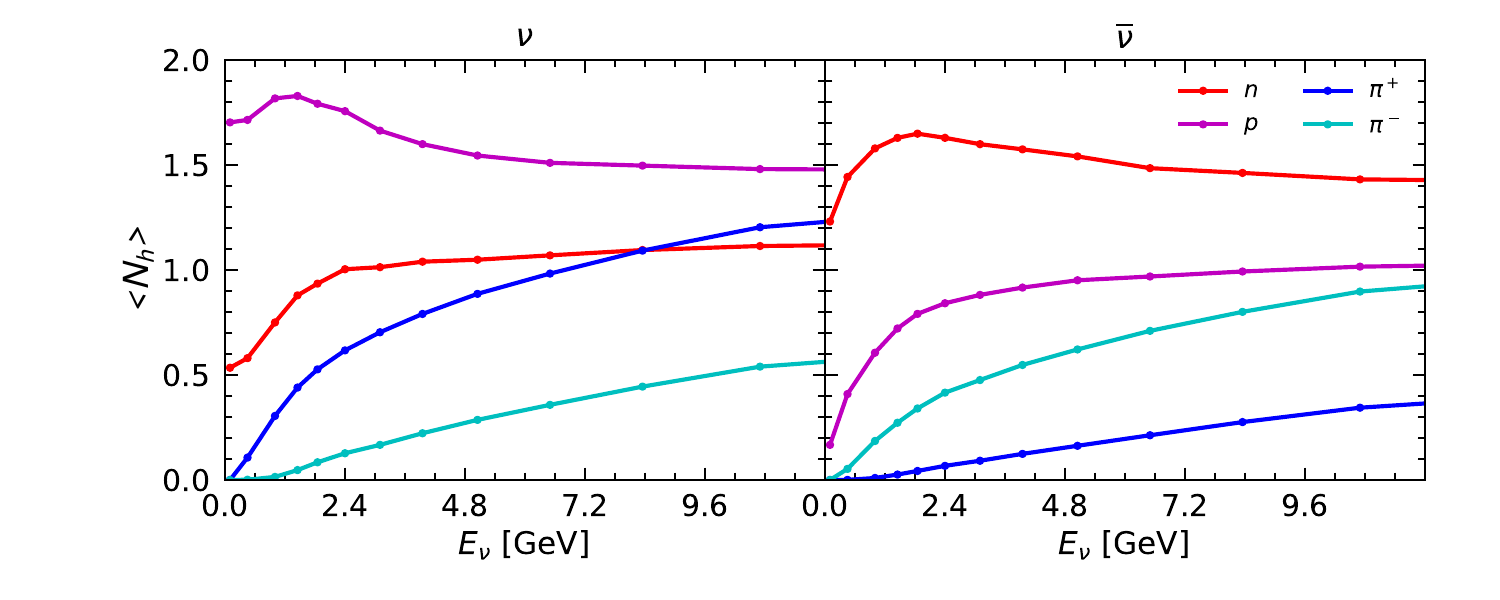}
        \caption{The {averaged} multiplicity of final-state hadrons {$\langle N_{h}\rangle$} as a function of the incoming neutrino energy for neutrinos (left) and antineutrinos (right). The {solid lines with dots} represent the mean multiplicity in each energy interval.}
        \label{fig:Enu_FSparticles_profile}
    \end{figure}

\begin{figure}
        \centering
        \includegraphics[width=0.95\textwidth]{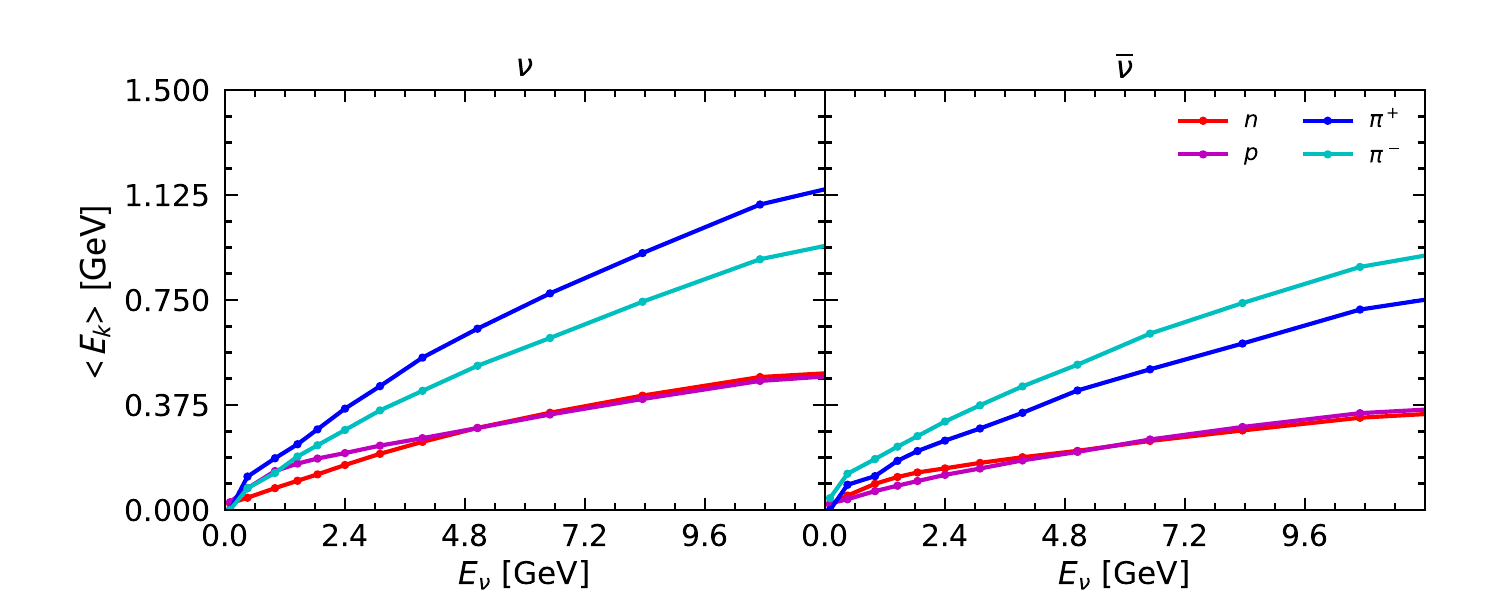}
        \caption{The {averaged} kinetic energy of the four primary final-state hadrons {$\langle E_k\rangle$} as a function of neutrino energy for neutrinos (left) and antineutrinos (right). The {solid lines with dots} represent the mean kinetic energy of the hadrons in each energy interval.}
        \label{fig:Enu_FSparticlesEk_profile}
    \end{figure}

The primary interactions of atmospheric neutrinos in LS produce hadronic final states, which mainly consist of neutrons, protons, and $\pi^{\pm}$ mesons. Figure~\ref{fig:Enu_FSparticles_profile} illustrates the average multiplicity of these final-state hadrons as a function of the neutrino energy. One can observe that antineutrino interactions generate more neutrons and $\pi^{-}$'s than neutrino interactions, whereas neutrino interactions yield a higher number of protons and $\pi^{+}$'s. Additionally, within the same hadronic isospin multiplet—such as protons and neutrons—there is also an observed asymmetry between neutrinos and antineutrinos. 
At the same incident energies, neutrinos generate a greater number of protons compared to the number of neutrons produced by antineutrinos. As previously discussed, neutrinos transfer more energy to the final-state hadronic system compared to antineutrinos, leading to generally higher multiplicities of final-state hadrons for neutrinos.
In Figure~\ref{fig:Enu_FSparticlesEk_profile}, the average kinetic energies of neutrons, protons, and $\pi^{\pm}$ mesons as a function of neutrino energy are presented. Due to their lighter mass, $\pi^{\pm}$ mesons can acquire greater kinetic energy from interactions, while neutrons and protons exhibit relatively lower average kinetic energies. Furthermore, in the charged-current interactions, neutrinos transfer more energy to the hadronic system compared to antineutrinos. As a result, the average kinetic energies of all four types of hadrons are significantly higher in neutrino-induced events than in antineutrino-induced events.

    \begin{figure}
        \centering
        \includegraphics[width=0.8\textwidth]{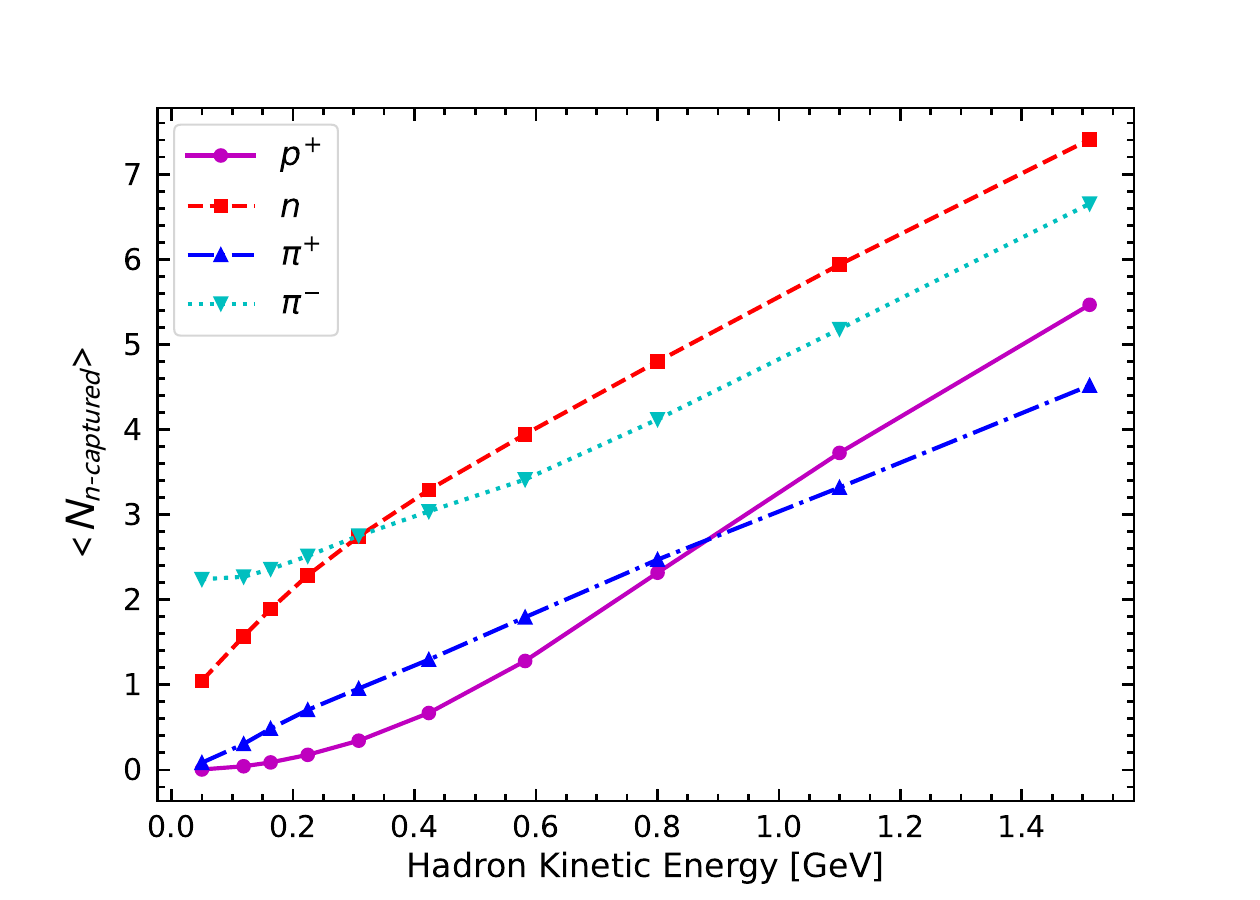}
        \caption{The {averaged} captured neutron multiplicity {$\langle N_{\rm n-captured}\rangle$} as a function of kinetic energy for the four primary final-state hadrons. The {lines with dots} represent the mean value of captured neutron multiplicity in each energy interval.}
        \label{fig:hadronEk_nMutli}
    \end{figure}

Neutron capture is commonly used for detecting low-energy antineutrinos through the inverse beta decay process with free protons. However, the situation becomes significantly more complex when dealing with higher-energy neutrinos and bound nuclei. The primary interaction begins when a neutrino strikes a nucleus, triggering final-state interactions within the nucleus that can produce neutrons. In this mechanism, antineutrinos are more efficient than neutrinos at generating neutrons. Furthermore, hadrons produced in the primary interactions can induce secondary interactions in the LS, such as inelastic scattering and $\pi^{-}$ capture, which further contribute to neutron production. Finally, all resulting neutrons are captured by nuclei, predominantly by protons in LS, emitting characteristic gamma-ray signals. Understanding this complex interplay between neutrino interactions, neutron production in LS, and subsequent neutron captures is essential for advancing our knowledge of the captured neutron multiplicity distribution.

To elucidate the physics underlying the neutron capture multiplicity distributions, we provide a detailed analysis of their production and interaction processes. First, the number of neutrons produced in hadronic secondary interactions is strongly influenced by the kinetic energy of the outgoing hadrons---primarily protons, neutrons, and $\pi^{\pm}$. Figure~\ref{fig:hadronEk_nMutli} shows the mean captured neutron multiplicity as a function of kinetic energy for four final-state hadrons. As kinetic energy increases, the neutron production capability of these hadrons becomes more pronounced. Notably, both neutrons and $\pi^{-}$ exhibit particularly high neutron yields. The $\pi^{-}$ is of special interest because it can not only induce secondary interactions that produce additional neutrons but also be captured by nuclei, further enhancing neutron generation.

   \begin{figure}
        \centering
        \includegraphics[width=0.92\textwidth]{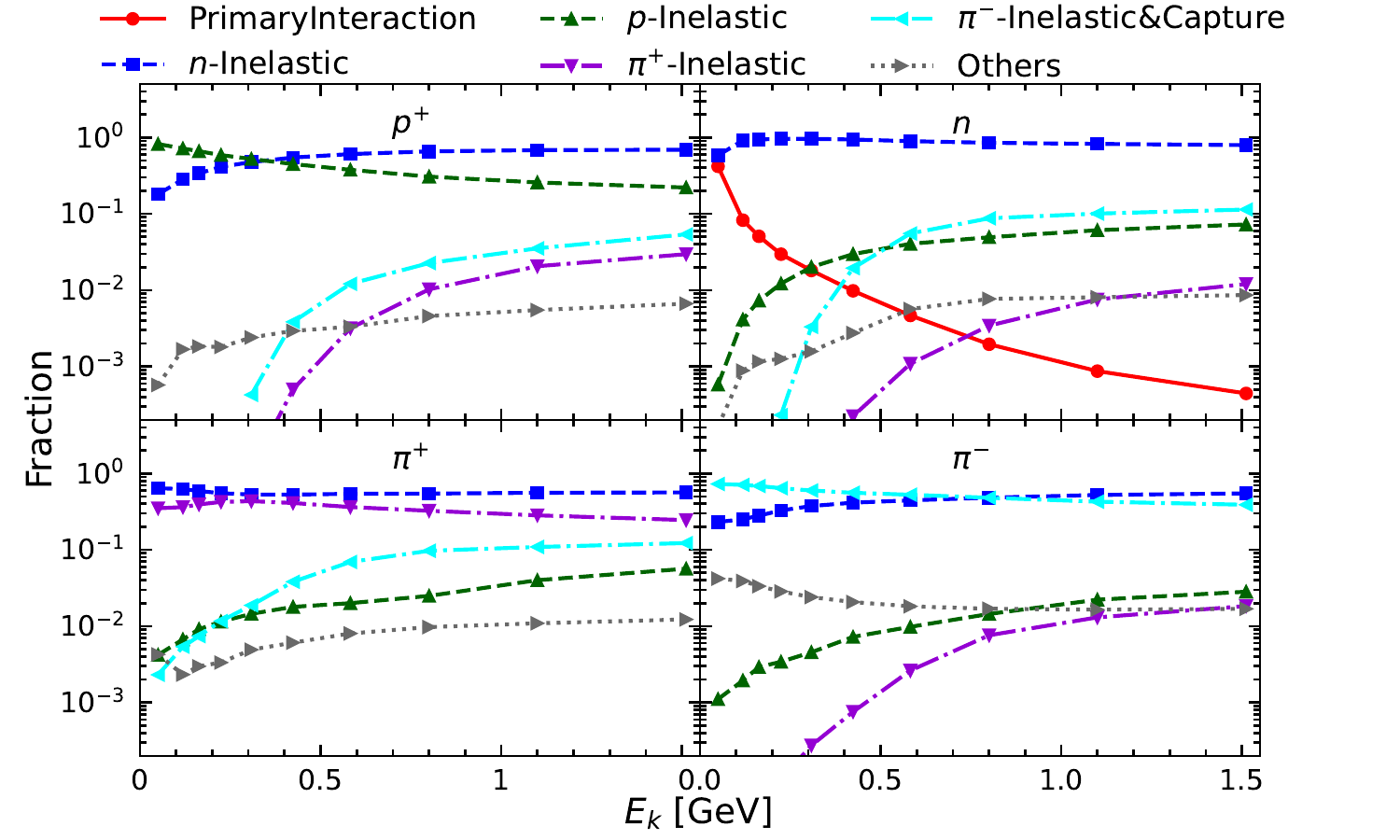}
        \caption{The fractions of different neutron production processes for four final-state hadrons as functions of their kinetic energies. The fraction is defined as the number of captured neutrons produced by a specific process divided by the total number of captured neutrons. All of these neutrons are counted separately for protons, neutrons, $\pi^{+}$'s, and $\pi^{-}$'s generated from primary neutrino interactions. The x-axis represents the kinetic energy of these four types of final-state hadrons.}
        \label{fig:neutron_production}
    \end{figure}

To further clarify the production mechanisms of captured neutrons, Figure~\ref{fig:neutron_production} shows the relative contributions of different neutron production processes for each of the four final-state hadrons as a function of kinetic energy. Neutron inelastic scattering is observed to be a dominant source of neutron production across all final-state hadrons. Besides neutron inelastic scattering, inelastic interactions induced by protons and charged pions also contribute substantially to neutron yields. Subdominant contributions come from photonuclear interactions and decays of unstable hadrons and nuclei. The neutrons generated from primary neutrino interactions account for tens of percent of neutron captures at kinetic energies below 100 MeV, but their contribution declines rapidly with increasing energy.
    
    \begin{figure}
        \centering
        \includegraphics[width=0.92\textwidth]{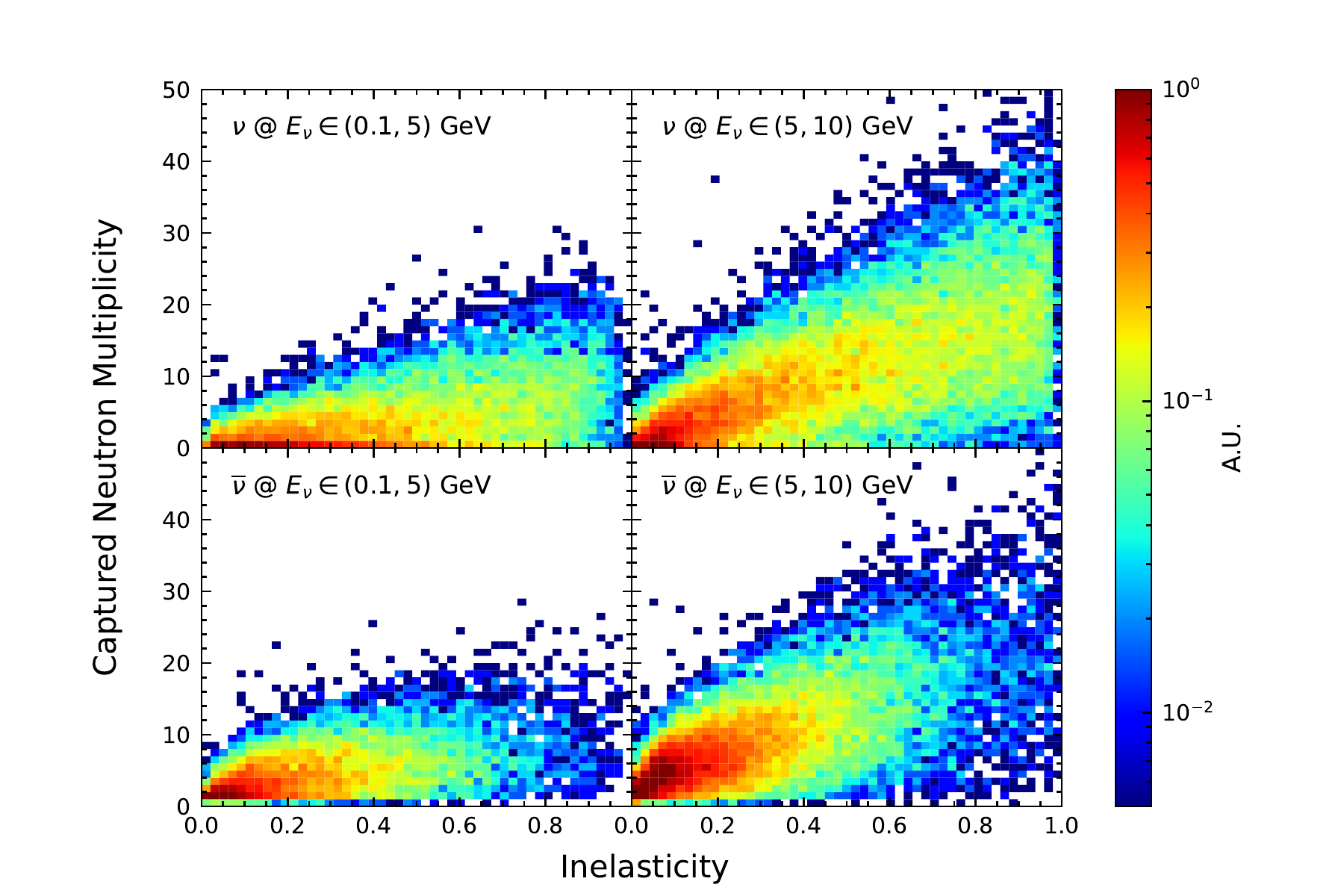}
        \caption{Two-dimensional distributions of inelasticity verus captured neutron multiplicity for neutrinos (upper panels) and antineutrinos (lower panels). The left and right panels correspond to neutrino/anineutrino energies below and above 5\,GeV, respectively. For each energy range, flat neutrino energy spectra are utilized.}
        \label{fig:y_nMutli_hist2d}
    \end{figure}

\begin{figure}
        \centering
        \includegraphics[width=0.95\textwidth]{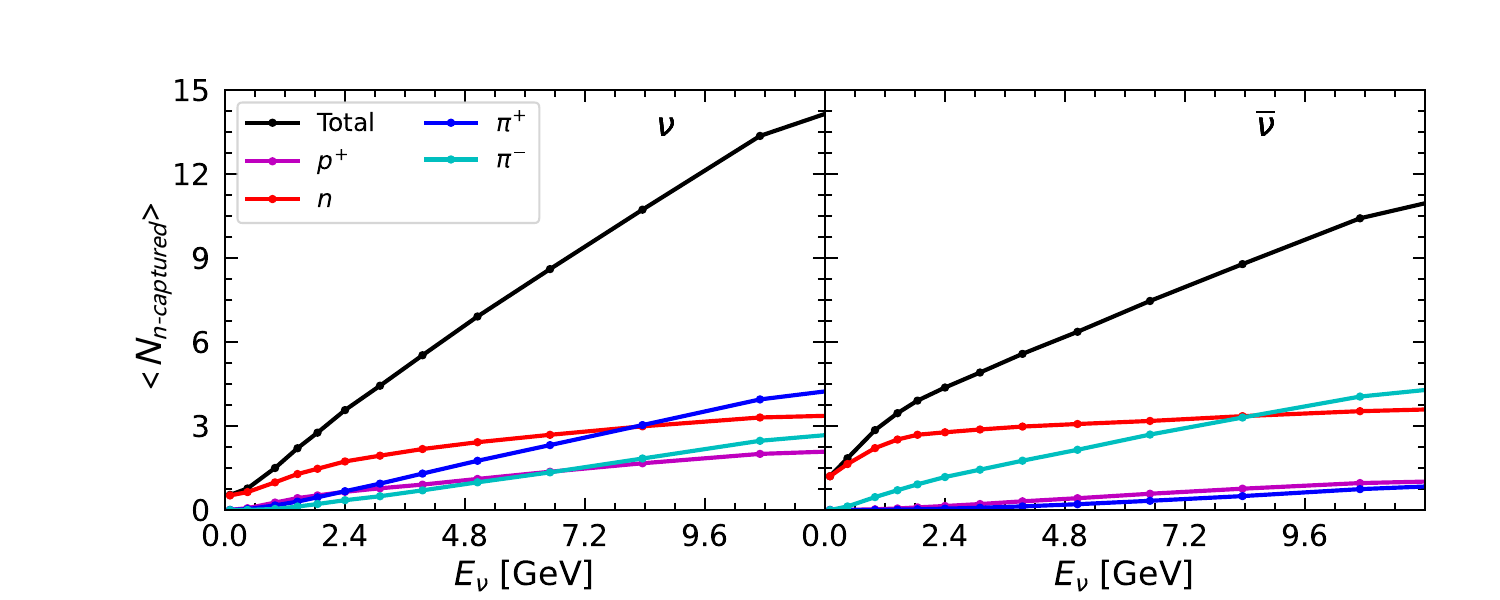}
        \caption{
        The averaged captured neutron multiplicity {$\langle N_{\rm n-captured}\rangle$} as functions of the neutrino (left panel) or antineutrino (right panel) energies is provided, the individual contributions of different final-state hadrons are also illustrated for comparison.}
        \label{fig:neutron_multiplicity}
    \end{figure}

As previously noted, neutrinos and antineutrinos exhibit distinct inelasticity distributions,
reflecting differences in the energy transferred to the hadronic system in each case. Given that the captured neutron multiplicity is closely related to hadron kinetic energies, both inelasticity and neutron multiplicity serve as correlated observables useful for distinguishing between neutrinos and antineutrinos. This relationship is illustrated in Figure~\ref{fig:y_nMutli_hist2d}, which displays two-dimensional distributions of inelasticity versus captured neutron multiplicity for neutrinos (upper panels) and antineutrinos (lower panels), separately for energies below (left) and above (right) 5 GeV. A clear positive correlation emerges above 5 GeV.
Compared to neutrinos, antineutrino interactions generally transfer less energy to the hadronic system but produce more neutrons through primary processes. Consequently, at lower energies antineutrino interactions yield more captured neutrons, while at higher energies neutrino interactions dominate in neutron yield. This energy-dependent behavior leads to clearly separable two-dimensional patterns in inelasticity and neutron multiplicity, providing a key handle for neutrino–antineutrino discrimination. The same feature is further illustrated in Figure~\ref{fig:neutron_multiplicity}, which shows the average captured neutron multiplicity as a function of energy for neutrinos (left) and antineutrinos (right), with individual contributions from different final-state hadrons indicated for comparison. At low energies, neutron inelastic processes dominate, with primary neutrons contributing more strongly in antineutrino interactions. At higher energies, charged-pion-induced processes become the most significant source of neutron production.
    
\section{Detector Size Effects}
\label{sec:geometry}

Based on the preceding discussions, the features of charged leptons, inelasticity, and captured neutron multiplicity have mainly been examined in terms of ideal physical effects. In practice, however, realistic detector configurations can significantly alter these distributions and impact the discrimination between neutrino types. Taking the JUNO detector as an example, which uses a 17.7\,m radius acrylic sphere containing 20\,kton LS as its target, we concentrate on fully-contained (FC) events, defined as interactions that deposit all their energy inside the target. This selection allows the study of the relevant observables in a well-contained and controlled environment. In contrast, partially-contained (PC) events are those that begin within the central detector (CD) but have a portion of their energy outside the CD region.

For electron neutrino interactions, the spatial extent of the final-state electron is generally small because the electromagnetic shower deposits energy rapidly. As a result, most of these events are FC, and their physical properties are only weakly affected by the detector size. The situation is notably different for muon neutrino interactions. As outlined in Section~\ref{section:lepton}, the final-state muon usually produces an extended track inside the detector. Hence, FC muon neutrino events are strongly influenced by geometric constraints, which can alter their intrinsic characteristics including energy measurement, track reconstruction, and the ability to distinguish between neutrino types.

    \begin{figure}
        \centering
        \includegraphics[width=0.95\textwidth]{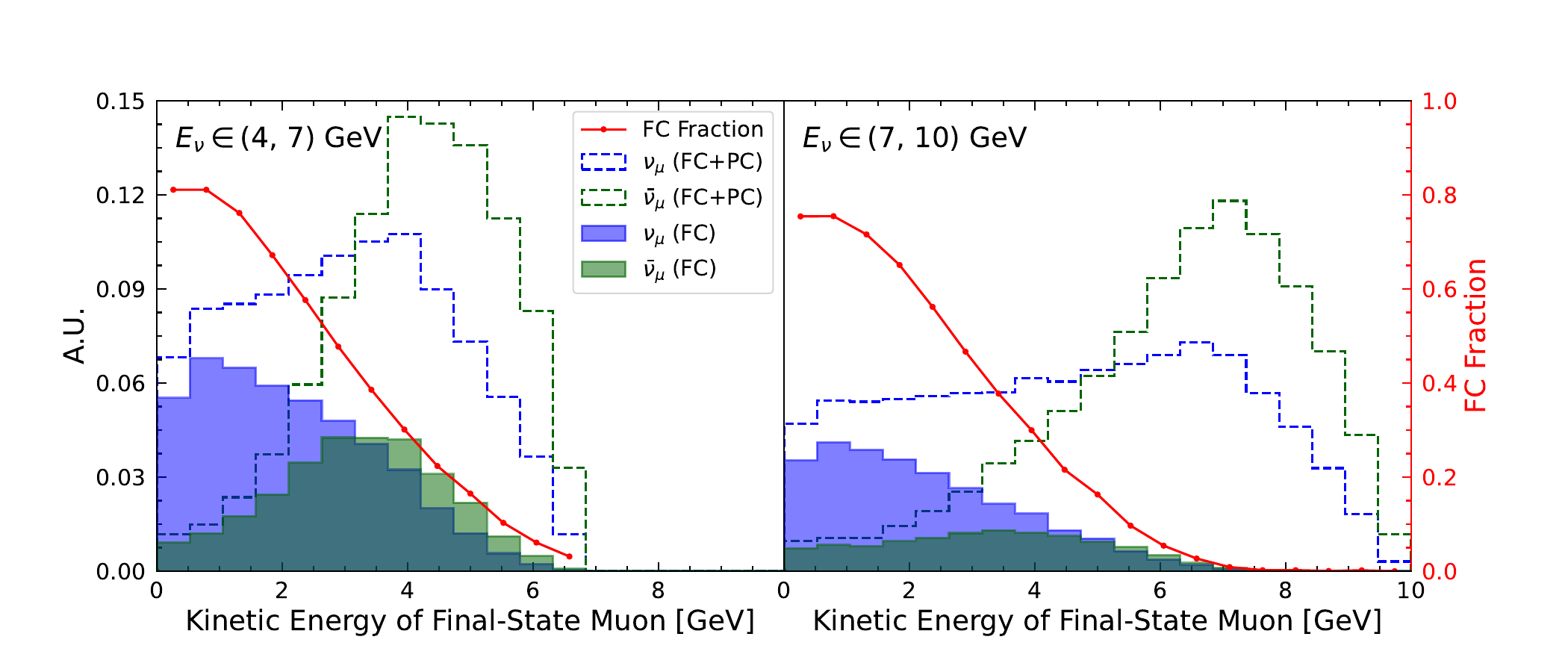}
        \caption{
        The influence of the detector size on the kinetic energy distributions of muon neutrino samples.
        Dashed lines represent the combined distributions from all FC and PC events; solid lines correspond to FC events only. Blue and green lines represent neutrino and antineutrino samples, respectively. The red solid lines indicate the FC fraction as a function of kinetic energy, which is identical for both \(\nu_{\mu}\) and \(\overline{\nu}_{\mu}\) events. The left and right panels correspond to different incident neutrino energy ranges, which are simulated by using flat neutrino energy spectra.}
        \label{fig:lep_Ek_1d}
    \end{figure}

The finite detector size distorts the measured kinetic energy spectra of muons due to the physical cutoff on observable muon track lengths, especially at high energies. This geometric effect is analyzed in Figure~\ref{fig:lep_Ek_1d}, which compares the spectra from neutrino and antineutrino interactions under different containment conditions. The fraction of events that are fully contained within the active volume serves as a direct measure of geometric acceptance; this fraction decreases as the muon energy increases. Consequently, although antineutrinos intrinsically produce higher-energy muons because they transfer less energy to hadrons, this difference is notably reduced in the spectrum of fully-contained events. At high incident energies, the geometric cutoff strongly suppresses the high-energy tail of the antineutrino spectrum, leading to a convergence of the observed distributions for the two flavors.

\begin{figure}
        \centering
        \subfigure[]{
            \includegraphics[width=0.48\textwidth]{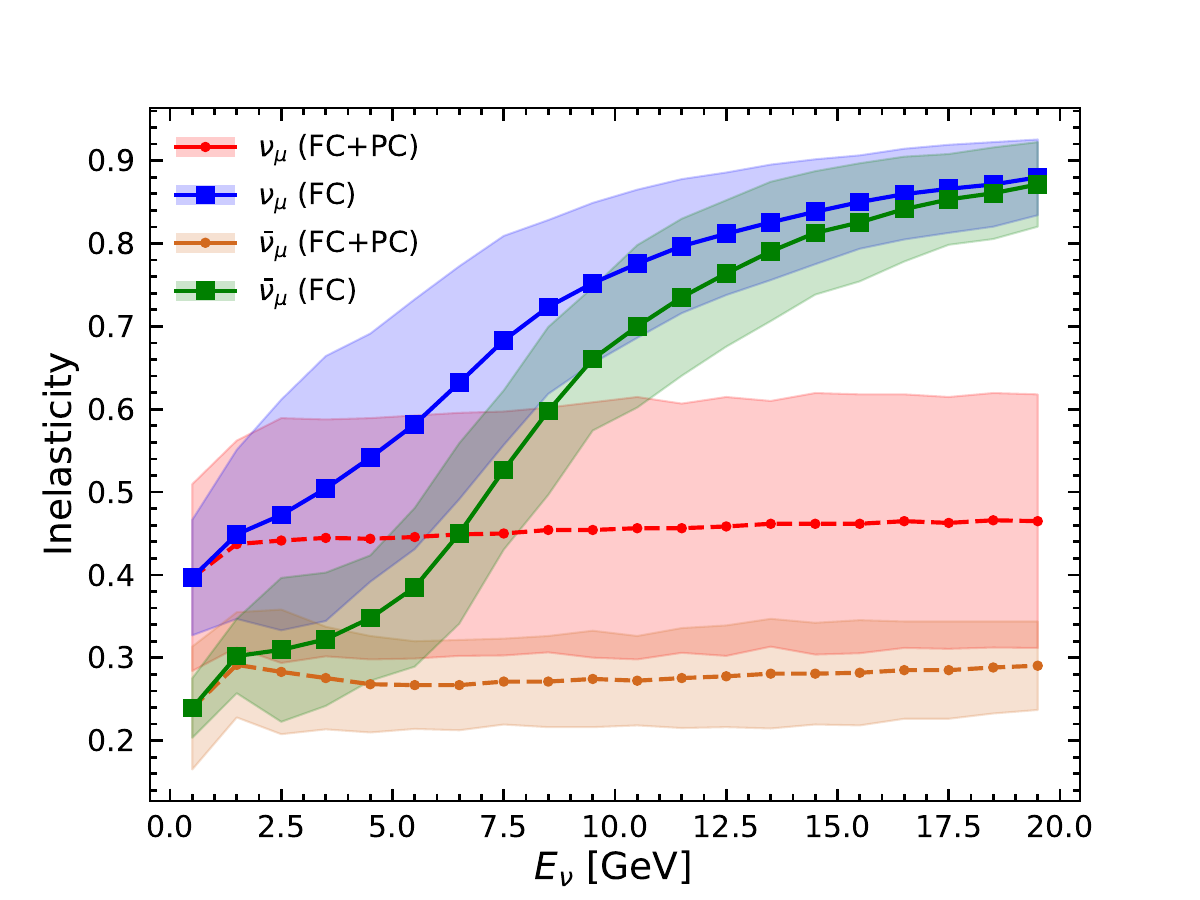}
            \label{fig:y_ratio}
        }
        \hfill
    	\subfigure[]{
        	\includegraphics[width=0.48\textwidth]{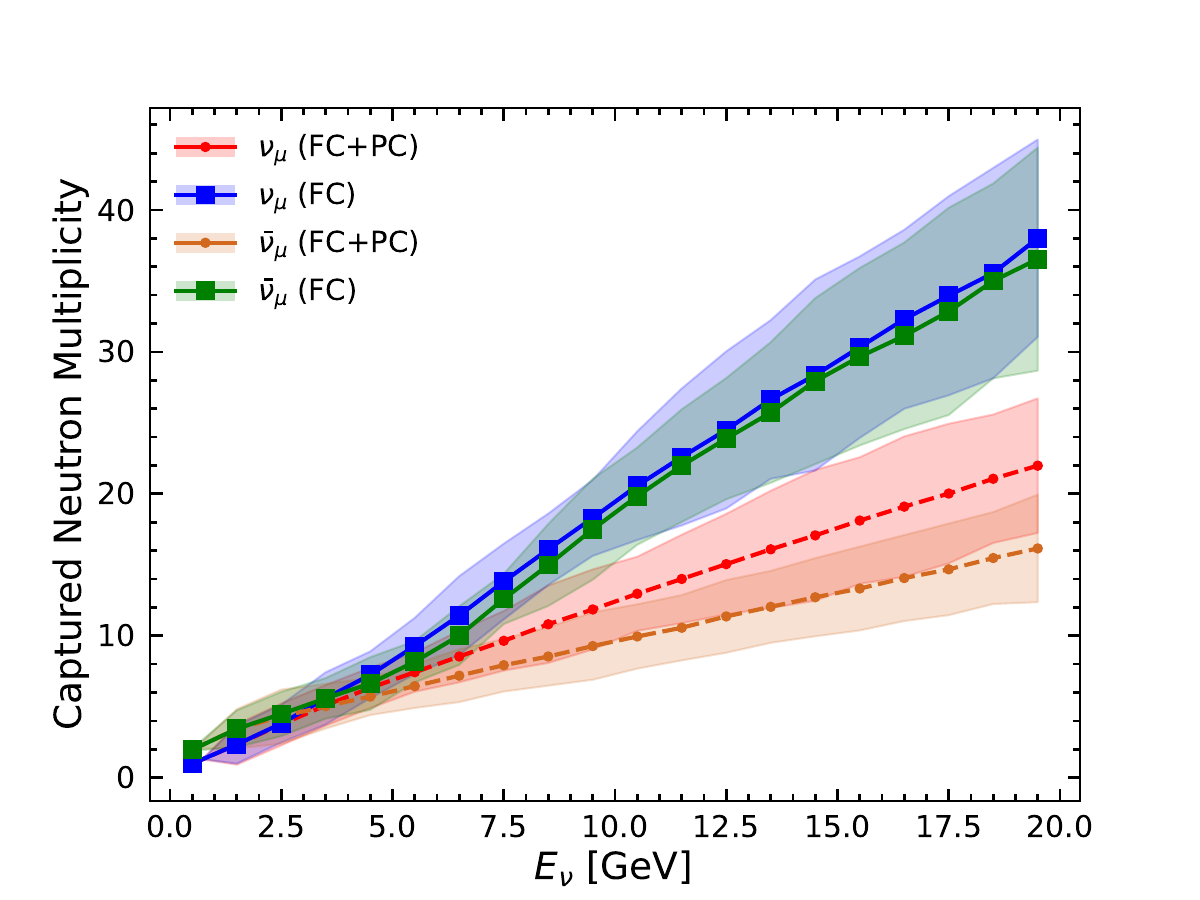}
        	\label{fig:nMulti}
    	}
        \caption{{The inelasticity (left panel) and captured neutron multiplicity (right panel) distributions as functions of the muon neutrino energy. The dashed lines with dots represent the the mean values for all FC and PC events, while the solid lines with dots indicate the mean values for FC events only. The shaded bands indicate the standard deviation of distributions within each energy interval.}}
        \label{fig:features_FC}
    \end{figure}

Building on the geometric distortion of muon energy, the inelasticity and captured-neutron multiplicity in fully-contained (FC) $\nu_\mu$/$\overline{\nu}_\mu$ events are also reshaped by the detector boundary, as shown in Figure~\ref{fig:features_FC}. At high neutrino energies, the detectable muon energy is limited by the detector boundary, forcing a larger share of the energy to be assigned to the hadronic system in {the} reconstruction. 
This artificially inflates both the measured inelasticity and the neutron multiplicity, while simultaneously suppressing the genuine flavor differences that depend on the detailed energy partition, making $\nu_\mu$/$\overline{\nu}_\mu$ separation difficult for high-energy FC events. 
{In Figure~\ref{fig:features_FC}, besides the mean values depicted as solid and dashed lines with dots, the corresponding bands are also illustrated as the standard deviations of the distributions within each energy interval, which serve as an effective resolution arising from both the neutrino interaction and secondary interaction.}
This effect is inherently tied to the detector’s physical scale. Given JUNO’s liquid-scintillator diameter of 35.4\,m, muons with kinetic energies above $\sim$7\,GeV typically escape the active volume. Hence, {the} majority of the events beyond this threshold are necessarily partially contained. Conversely, in the few-GeV energy range, which is most relevant for the NMO determination, the detector size effect turns from a limitation into an advantage. {In this regime}, muons are fully contained, and the hadronic system receives enough energy to amplify flavor-sensitive signatures in inelasticity and neutron multiplicity. This provides a clear window for effective $\nu_\mu$/$\overline{\nu}_\mu$ discrimination,  precisely where it matters for the NMO sensitivity.

\section{Neutrino-Antineutrino Discrimination}
\label{sec:separation}

In this section, we quantitatively assess the performance of neutrino-antineutrino discrimination based on the event-level observables discussed previously. As an illustration, we adopt a simple BDT model that uses inelasticity and captured neutron multiplicity as discriminating variables. More advanced machine-learning techniques using the PMT level information have been explored for flavor identification elsewhere (see details in Ref.~\cite{Liu:2025fry}) and are not considered here.

    \begin{figure}
        \centering
        \includegraphics[width=0.8\textwidth]{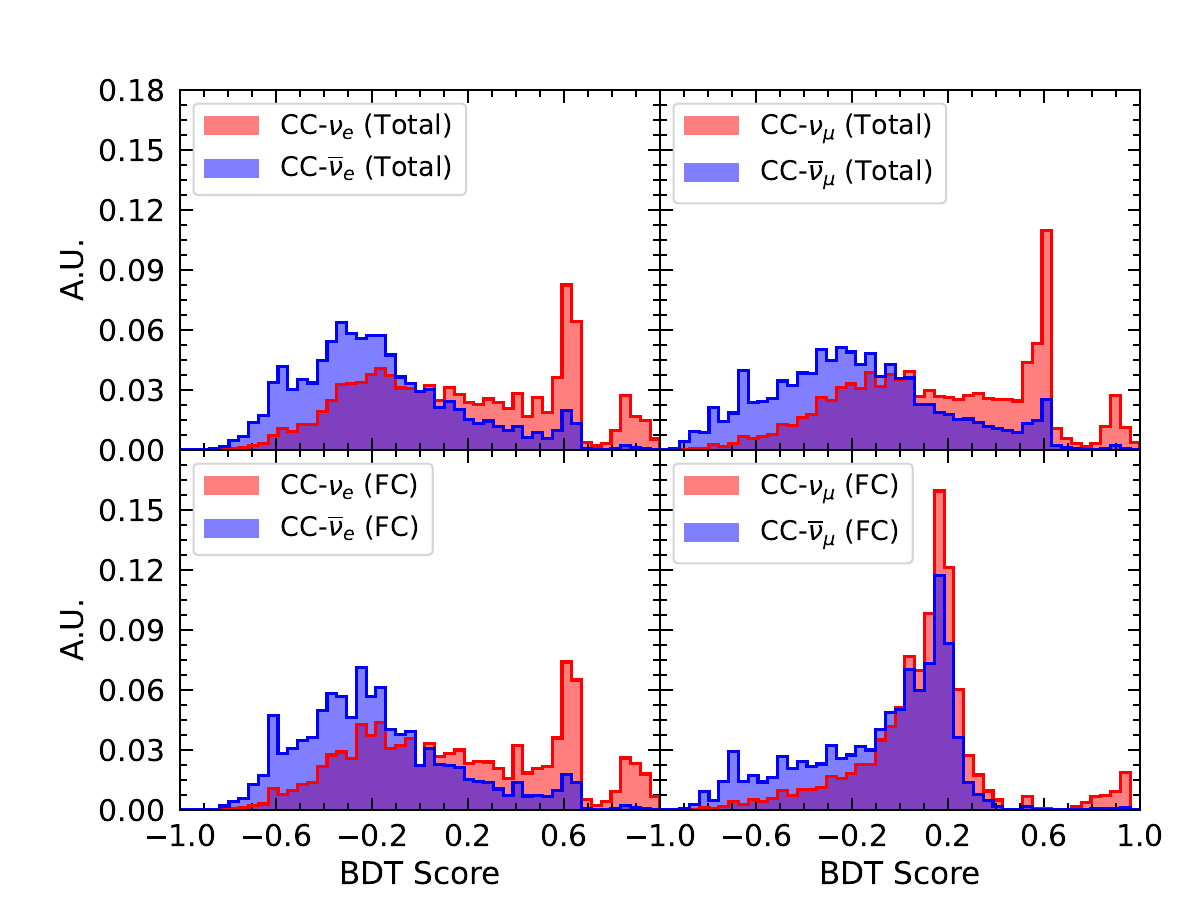}
        \caption{The BDT score distribution of $\nu_{e}$/$\overline{\nu}_{e}$ (left) and $\nu_{\mu}$/$\overline{\nu}_{\mu}$ (right) discrimination in the cases of total samples (top) and FC samples (bottom).}
        \label{fig:BDT_score}
    \end{figure}
    \begin{figure}[h!]
        \centering
        \includegraphics[width=0.8\textwidth]{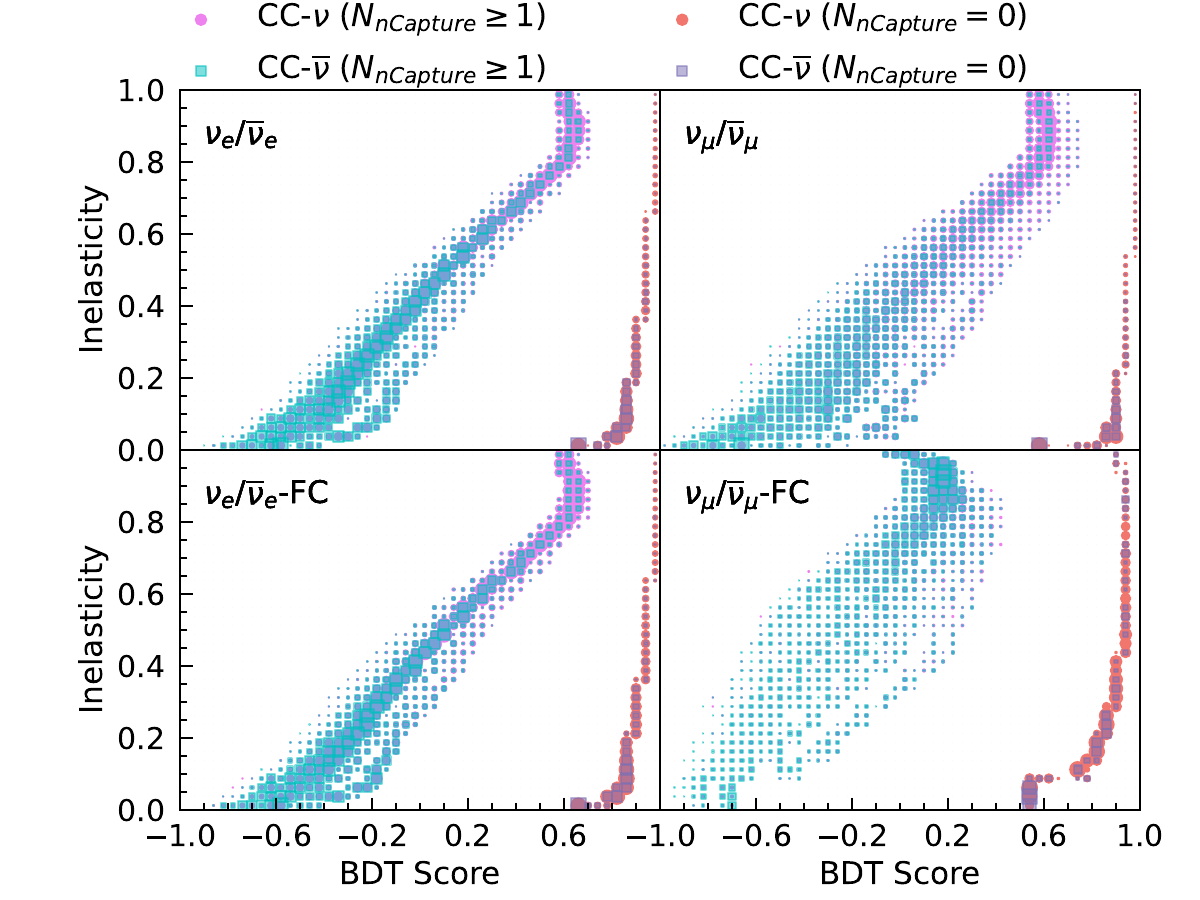}
        \caption{The distribution of inelasticity and BDT score in the cases of with and without the neutron captures. The left and right panels are for 
        $\nu_{e}$/$\overline{\nu}_{e}$ and $\nu_{\mu}$/$\overline{\nu}_{\mu}$ discrimination respectively, while the upper and lower panels are for the total and FC samples.}
        \label{fig:score_Y_ratio}
    \end{figure}

    \begin{figure}
        \centering
        \subfigure[]{
            \includegraphics[width=0.48\textwidth]{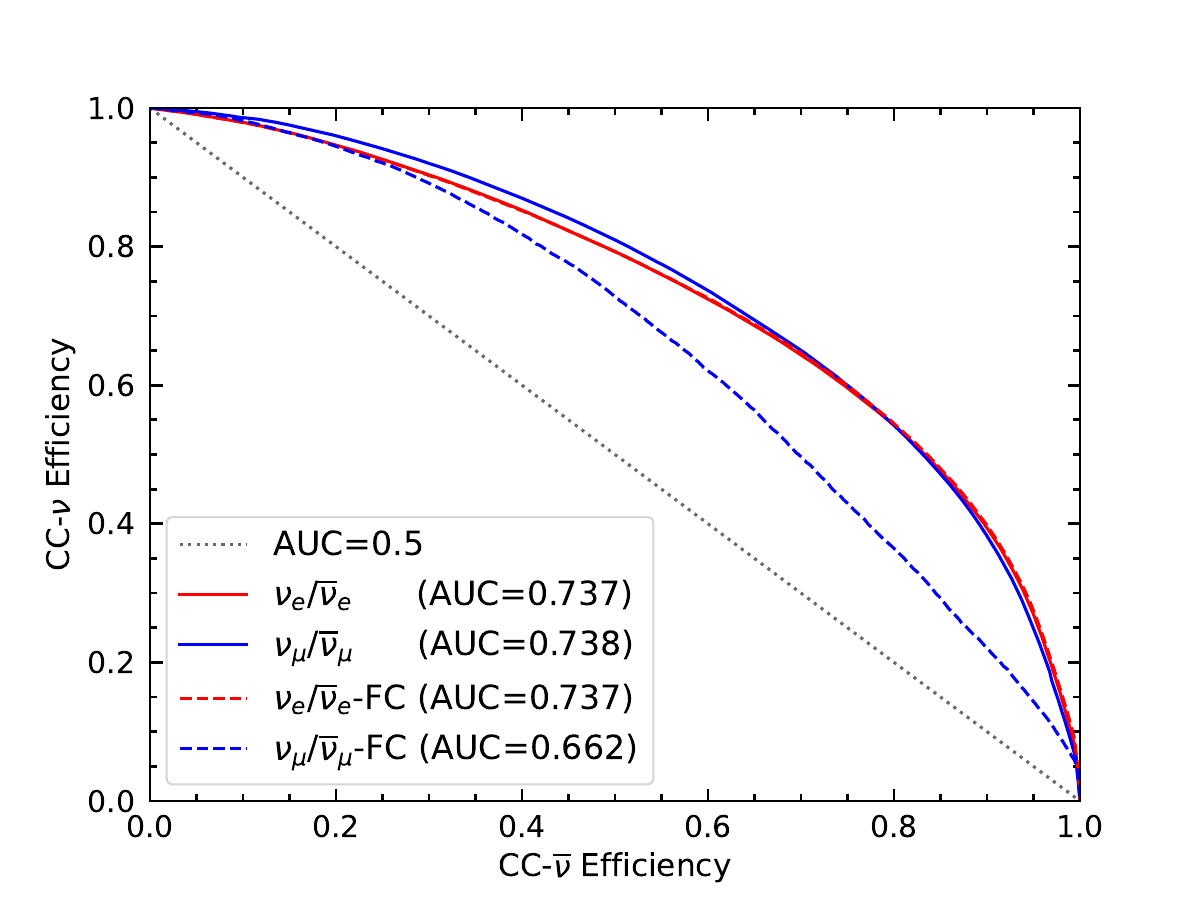}
            \label{fig:ROC_curve}
        }
        \hfill
        \subfigure[]{
            \includegraphics[width=0.48\textwidth]{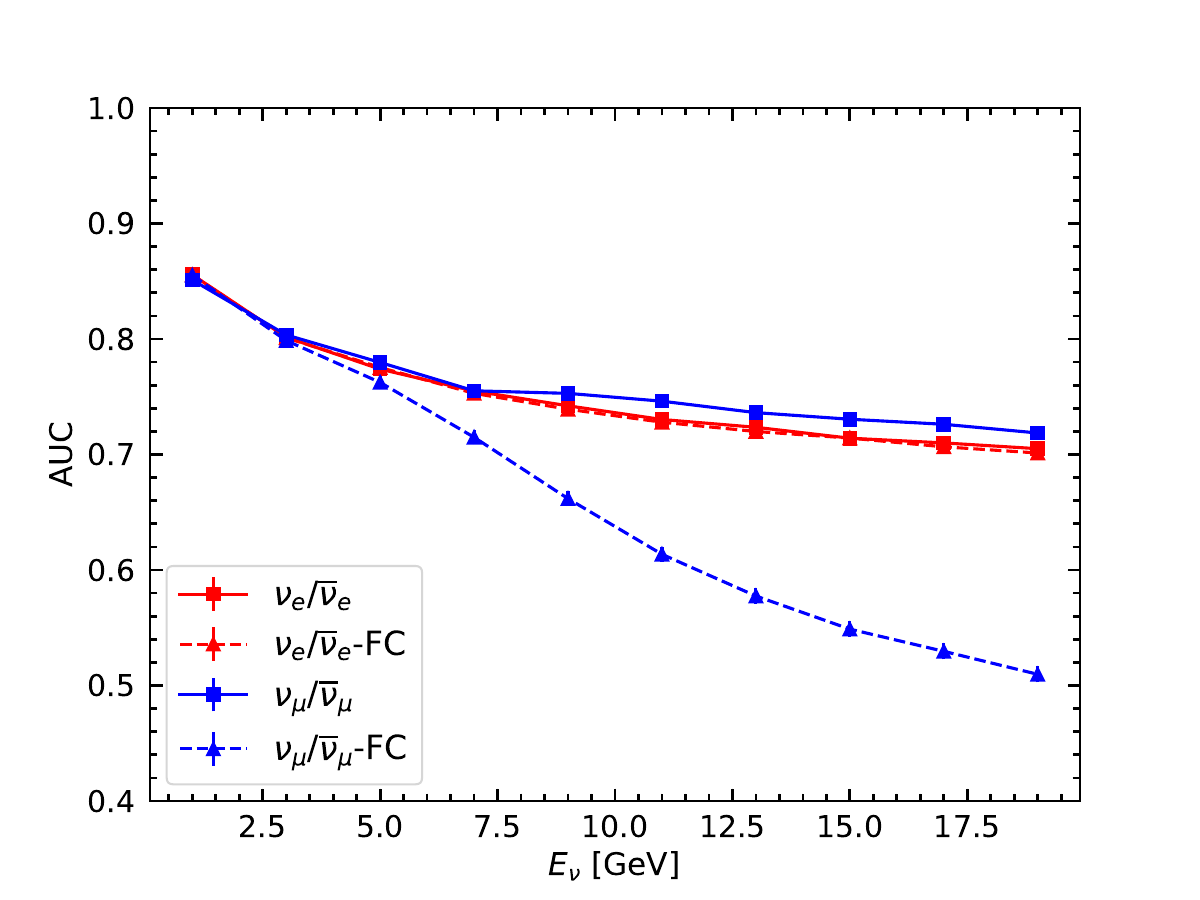}
            \label{fig:AUC_vs_Enu}
        }
        \caption{The ROC curve (left panel) and AUC values (right panel) for the neutrino-antineutrino discrimination.
        The discrimination of $\nu_{e}$/$\overline{\nu}_{e}$ and $\nu_{\mu}$/$\overline{\nu}_{\mu}$ for the total and FC samples are shown separately using the true values of inelasticity and neutron multiplicity obtained from simulation}. The entire energy range has been integrated for the left panel while displayed for the right panel.
        \label{fig:BDT_performace}
    \end{figure}

We have trained BDT classifiers separately for different flavors to evaluate their discrimination power between neutrinos and antineutrinos {in the energy range from 0.1 to 20\,GeV}, using the area under the curve (AUC) of the Receiver Operating Characteristic (ROC) curve as the performance metric.
In Figure~\ref{fig:BDT_score}, we illustrate the BDT score distributions of $\nu_{e}$/$\overline{\nu}_{e}$ (left) and $\nu_{\mu}$/$\overline{\nu}_{\mu}$ (right) discrimination in the cases of total samples (top) and FC samples (bottom). The higher the BDT score, the more an event type resembles the neutrino category as opposed to the antineutrino category. From the figure one {finds} that the BDT scores are smaller for antineutrino samples {(blue curves)}, compared to those of neutrino samples. In addition, the red curves of neutrino samples exhibit two peaks at $\sim$0.6 and $\sim$0.9. To understand these specific peak features, we show in Figure~\ref{fig:score_Y_ratio} the two-dimensional distributions of inelasticity and BDT score {for cases with or without a true neutron capture}. 
As illustrated in the figure, the criterion for determining whether the captured neutron multiplicity is zero is a crucial factor for distinguishing between neutrino and antineutrino events. Samples without {neutron capture} show the highest BDT scores, peaking at 0.9. 
For samples with captured neutron production, higher inelasticity correlates with a higher BDT score. However, this relationship saturates when the inelasticity value exceeds 0.8, resulting in a plateau where the BDT score remains unchanged, with a peak of the BDT score observed at 0.6. This behavior can be understood through the inelasticity curves shown in Figure~\ref{fig:inelasticity}, where the decreasing inelasticity dependence of antineutrino DIS channels transitions to relatively flat curves.

In Figure~\ref{fig:BDT_performace}, we illustrate the ROC curve (left panel) and AUC values (right panel) for the neutrino-antineutrino discrimination. 
Figure~\ref{fig:ROC_curve} presents the ROC curves on the $\nu_{e}$/$\overline{\nu}_{e}$ and $\nu_{\mu}$/$\overline{\nu}_{\mu}$ discrimination for the total and FC samples. 
Figure~\ref{fig:AUC_vs_Enu} presents the AUC values as a function of neutrino energy, comparing results for both electron and muon flavors under different geometric containment conditions. The combined use of inelasticity and neutron multiplicity provides effective discrimination, which can {exceed} 80\% at lower energies when the neutron multiplicity contributes significant additional separating power. As energy increases, the discrimination power decreases; however, the AUC remains around 75\% in the energy range of approximately 3 to 10\,GeV (the NMO sensitive region).
For muon neutrinos in the FC sample, discrimination performance degrades at higher energies. This decline is a direct geometric effect: as the neutrino energy increases, the muon track length may exceed the detector’s active volume. This containment limit suppresses the measurable energy transfer to the final-state lepton and reduces {neutrino-type-sensitive} information in the hadronic system. Therefore, a detector with a larger active volume not only improves event statistics but also better preserves the intrinsic separation power, particularly for high-energy muon neutrinos, by allowing longer tracks to remain fully contained.

It is important to note that these results are obtained under idealized assumptions, without accounting for reconstruction resolution in inelasticity or tagging efficiency for captured neutrons. In LS detectors, the low energy threshold significantly enhances the detection capability for hadronic components, including signals from neutron capture. 
This capability not only improves neutrino–antineutrino discrimination but also enables deeper insights into the underlying physics of atmospheric neutrino interactions in {the} LS medium.

Before concluding this work, we highlight the neutrino-antineutrino discrimination potential across different detector technologies.
The Super-Kamiokande experiment has been upgraded with gadolinium loading since 2020 to improve neutron tagging efficiency, enhancing its sensitivity to the diffuse supernova neutrino background (DSNB)~\cite{Super-Kamiokande:2021the} and benefiting atmospheric neutrino oscillation studies~\cite{Super-Kamiokande:2023ahc}. Recent {studies} have also explored the use of final-state hadron detection and machine learning to reconstruct inelasticity in water/ice Cherenkov detectors such as IceCube-Upgrade and ORCA~\cite{IceCube:2018pgc, Olavarrieta:2024eaq}. The upcoming DUNE experiment will employ liquid-argon time-projection chamber technology~\cite{DUNE:2020mra}, offering excellent tracking and calorimetric precision for both charged leptons and hadrons from neutrino–nucleus charged-current interactions~\cite{Friedland:2018vry}. However, neutron tagging remains challenging in such detectors, which may affect energy resolution and limit the potential for distinguishing neutrinos from antineutrinos.

\section{Conclusion}
\label{sec:conclusion}

In this work, we have systematically evaluated the characteristics of charged-current interactions induced by atmospheric neutrinos on $^{12}{\rm C}$ and $^{1}{\rm H}$ nuclei in liquid-scintillator detectors, covering both the charged-lepton and hadronic components. The {principal} features of these event signatures can be summarized as follows.

\begin{itemize}
  \item {\bf Charged leptons and hadrons} exhibit distinct light topologies in LS: muons produce clear, track-like ionization patterns, electrons generate diffuse, multi-point electromagnetic showers, while hadrons yield multiple-component hadronic showers. This difference provides a strong basis for flavor and neutrino-antineutrino identification at the PMT level.
  \item {\bf Inelasticity}, defined as the fraction of energy transferred to the hadronic system, differs significantly between neutrinos and antineutrinos. Neutrinos produce a nearly flat inelasticity distribution, whereas antineutrinos preferentially populate events with smaller hadronic energy shares, resulting in a distribution that rapidly declines as the inelasticity value rises.
  \item {\bf Captured neutron multiplicity} is dominated by primary neutron production at low neutrino energies, but at higher energies secondary interactions of final-state hadrons in the LS medium become predominant. Below about 5\,GeV, antineutrino interactions yield higher neutron multiplicities; above this energy, the trend reverses, with neutrino interactions producing more neutrons. 
  \item 
  {\bf Detector size} significantly affects muon-flavor events due to their long tracks, while electron-flavor events remain largely unaffected. For contained muon events, both inelasticity and neutron multiplicity are enhanced by geometric containment, but the intrinsic differences between neutrinos and antineutrinos are simultaneously suppressed.
\end{itemize}

By using the two-dimensional distribution of inelasticity and neutron multiplicity in a BDT classifier, we demonstrate a discrimination power between neutrinos and antineutrinos with AUC values exceeding 70\% under {an} idealized detection scenario. Performance is better at lower energies, where primary neutron production provides additional separating information. For fully-contained muon events, the AUC drops toward 50\% around 20\,GeV, a direct consequence of geometric containment that limits measurable lepton energy and suppresses flavor-sensitive hadronic information.
Further studies utilizing more advanced machine learning methods that take into account additional features at the PMT level are anticipated to yield improved performance.

The present analysis also enables improved energy and direction reconstruction for atmospheric neutrino charged-current events in LS detectors~\cite{Yang:2023rbg}. 
Collectively, these results establish a physics foundation for advanced high-energy neutrino reconstruction and flavor discrimination, which will enable more precise studies of neutrino oscillations, particularly in measurements of the neutrino mass ordering and CP violation with atmospheric neutrinos.
{
This work establishes the physical foundation and capability for neutrino/antineutrino discrimination in large LS detectors, utilizing truth-level quantities. However, the effects of detector response on inelasticity and neutron tagging cannot be neglected. For instance, the after-pulse effects~\cite{JUNO:2022hlz} that occur following a GeV-scale physical event can impact the tagging of captured neutrons produced by atmospheric neutrinos. Additionally, the resolution of inelasticity may introduce further confusion between neutrinos and antineutrinos. These effects necessitate further investigation using more accurate and comprehensive simulations, along with realistic data.
}

\section*{Acknowledgements}
The authors are grateful to Xianguo Lu and Jiajie Ling for helpful comments. 
{This work was supported in part by National Natural Science Foundation of China under Grant Nos. 12125506, 12075255, CAS Project for Young Scientists in Basic Research (YSBR-099)}

\bibliographystyle{h-physrev5}

\bibliography{references}

\begin{thebibliography}{10}

\bibitem{Kajita:2016cak}
T.~Kajita, {Nobel Lecture: Discovery of atmospheric neutrino oscillations},
\newblock Rev. Mod. Phys. {\bf 88}, 030501 (2016).

\bibitem{McDonald:2016ixn}
A.~B. McDonald, {Nobel Lecture: The Sudbury Neutrino Observatory: Observation
  of flavor change for solar neutrinos},
\newblock Rev. Mod. Phys. {\bf 88}, 030502 (2016).

\bibitem{Pontecorvo:1957cp}
B.~Pontecorvo, {Mesonium and Antimesonium},
\newblock Sov. Phys. JETP {\bf 6}, 429 (1958).

\bibitem{Maki:1962mu}
Z.~Maki, M.~Nakagawa, and S.~Sakata, {Remarks on the unified model of
  elementary particles},
\newblock Prog. Theor. Phys. {\bf 28}, 870 (1962).

\bibitem{Pontecorvo:1967fh}
B.~Pontecorvo, {Neutrino Experiments and the Problem of Conservation of
  Leptonic Charge},
\newblock Sov. Phys. JETP {\bf 26}, 984 (1968).

\bibitem{SNO:2011hxd}
SNO, B.~Aharmim {\em et~al.}, {Combined Analysis of all Three Phases of Solar
  Neutrino Data from the Sudbury Neutrino Observatory},
\newblock Phys. Rev. C {\bf 88}, 025501 (2013), arXiv:1109.0763.

\bibitem{T2K:2024wfn}
T2K, Super-Kamiokande, K.~Abe {\em et~al.}, {First Joint Oscillation Analysis
  of Super-Kamiokande Atmospheric and T2K Accelerator Neutrino Data},
\newblock Phys. Rev. Lett. {\bf 134}, 011801 (2025), arXiv:2405.12488.

\bibitem{Super-Kamiokande:2023ahc}
Super-Kamiokande, T.~Wester {\em et~al.}, {Atmospheric neutrino oscillation
  analysis with neutron tagging and an expanded fiducial volume in
  Super-Kamiokande I{\textendash}V},
\newblock Phys. Rev. D {\bf 109}, 072014 (2024), arXiv:2311.05105.

\bibitem{DayaBay:2012fng}
Daya Bay, F.~P. An {\em et~al.}, {Observation of electron-antineutrino
  disappearance at Daya Bay},
\newblock Phys. Rev. Lett. {\bf 108}, 171803 (2012), arXiv:1203.1669.

\bibitem{Wolfenstein:1977ue}
L.~Wolfenstein, {Neutrino Oscillations in Matter},
\newblock Phys. Rev. D {\bf 17}, 2369 (1978).

\bibitem{Mikheyev:1985zog}
S.~P. Mikheyev and A.~Y. Smirnov, {Resonance Amplification of Oscillations in
  Matter and Spectroscopy of Solar Neutrinos},
\newblock Sov. J. Nucl. Phys. {\bf 42}, 913 (1985).

\bibitem{DUNE:2021mtg}
DUNE, A.~Abud~Abed {\em et~al.}, {Low exposure long-baseline neutrino
  oscillation sensitivity of the DUNE experiment},
\newblock Phys. Rev. D {\bf 105}, 072006 (2022), arXiv:2109.01304.

\bibitem{Hyper-Kamiokande:2022smq}
Hyper-Kamiokande, J.~Bian {\em et~al.},
\newblock {Hyper-Kamiokande Experiment: A Snowmass White Paper},
\newblock in {\em {2022 Snowmass Summer Study}}, 2022, arXiv:2203.02029.

\bibitem{IceCube-PINGU:2014okk}
IceCube-PINGU, M.~G. Aartsen {\em et~al.}, {Letter of Intent: The Precision
  IceCube Next Generation Upgrade (PINGU)},
\newblock (2014), arXiv:1401.2046.

\bibitem{KM3NeT:2021ozk}
KM3NeT, S.~Aiello {\em et~al.}, {Determining the neutrino mass ordering and
  oscillation parameters with KM3NeT/ORCA},
\newblock Eur. Phys. J. C {\bf 82}, 26 (2022), arXiv:2103.09885.

\bibitem{JUNO:2015zny}
JUNO, F.~An {\em et~al.}, {Neutrino Physics with JUNO},
\newblock J. Phys. G {\bf 43}, 030401 (2016), arXiv:1507.05613.

\bibitem{JUNO:2024jaw}
JUNO, A.~Abusleme {\em et~al.}, {Potential to identify neutrino mass ordering
  with reactor antineutrinos at JUNO},
\newblock Chin. Phys. C {\bf 49}, 033104 (2025), arXiv:2405.18008.

\bibitem{Yang:2023rbg}
Z.~Yang {\em et~al.}, {First attempt of directionality reconstruction for
  atmospheric neutrinos in a large homogeneous liquid scintillator detector},
\newblock Phys. Rev. D {\bf 109}, 052005 (2024), arXiv:2310.06281.

\bibitem{Liu:2025fry}
J.~Liu {\em et~al.}, {Neutrino type identification for atmospheric neutrinos in
  a large homogeneous liquid scintillation detector},
\newblock Phys. Rev. D {\bf 112}, 012018 (2025), arXiv:2503.21353.

\bibitem{GENIE:2021zuu}
GENIE, J.~Tena-Vidal {\em et~al.}, {Neutrino-nucleon cross-section model tuning
  in GENIE v3},
\newblock Phys. Rev. D {\bf 104}, 072009 (2021), arXiv:2104.09179.

\bibitem{refId0}
{Ivanchenko, Vladimir} {\em et~al.}, Progress of geant4 electromagnetic physics
  developments and applications,
\newblock EPJ Web Conf. {\bf 214}, 02046 (2019).

\bibitem{Allison2016}
J.~A. et~al., Recent developments in geant4,
\newblock Nuclear Instruments and Methods in Physics Research A {\bf 835}, 186
  (2016).

\bibitem{Lin:2022htc}
T.~Lin {\em et~al.}, {Simulation Software of the JUNO Experiment},
\newblock (2022), arXiv:2212.10741.

\bibitem{Coppi:2023nlv}
A.~Coppi {\em et~al.}, {Mass testing of the JUNO experiment 20-inch PMT readout
  electronics},
\newblock Nucl. Instrum. Meth. A {\bf 1052}, 168255 (2023), arXiv:2301.04379.

\bibitem{IceCube:2018pgc}
IceCube, M.~G. Aartsen {\em et~al.}, {Measurements using the inelasticity
  distribution of multi-TeV neutrino interactions in IceCube},
\newblock Phys. Rev. D {\bf 99}, 032004 (2019), arXiv:1808.07629.

\bibitem{Ribordy:2013xea}
M.~Ribordy and A.~Y. Smirnov, {Improving the neutrino mass hierarchy
  identification with inelasticity measurement in PINGU and ORCA},
\newblock Phys. Rev. D {\bf 87}, 113007 (2013), arXiv:1303.0758.

\bibitem{Olavarrieta:2024eaq}
S.~G. Olavarrieta, M.~Jin, C.~A. Arg{\"u}elles, P.~Fern{\'a}ndez, and
  I.~Mart{\'\i}nez-Soler, {Boosting neutrino mass ordering sensitivity with
  inelasticity for atmospheric neutrino oscillation measurement},
\newblock Phys. Rev. D {\bf 110}, L051101 (2024), arXiv:2402.13308.

\bibitem{Super-Kamiokande:2021the}
Super-Kamiokande, K.~Abe {\em et~al.}, {First gadolinium loading to
  Super-Kamiokande},
\newblock Nucl. Instrum. Meth. A {\bf 1027}, 166248 (2022), arXiv:2109.00360.

\bibitem{DUNE:2020mra}
DUNE, B.~Abi {\em et~al.}, {Deep Underground Neutrino Experiment (DUNE), Far
  Detector Technical Design Report, Volume III: DUNE Far Detector Technical
  Coordination},
\newblock JINST {\bf 15}, T08009 (2020), arXiv:2002.03008.

\bibitem{Friedland:2018vry}
A.~Friedland and S.~W. Li, {Understanding the energy resolution of liquid argon
  neutrino detectors},
\newblock Phys. Rev. D {\bf 99}, 036009 (2019), arXiv:1811.06159.

\bibitem{JUNO:2022hlz}
JUNO, A.~Abusleme {\em et~al.}, {Mass testing and characterization of 20-inch
  PMTs for JUNO},
\newblock Eur. Phys. J. C {\bf 82}, 1168 (2022), arXiv:2205.08629.

\end{thebibliography}

\end{document}